\documentclass[twocolumn]{aastex63}

\usepackage{enumitem}
\usepackage{rotating}
\usepackage{threeparttable}
\usepackage{graphicx}
\usepackage{captcont}
\graphicspath{ {images/} }

\received{Month 00, 2020}
\revised{Month 00, 2020}
\accepted{Month 00, 2021} 
\submitjournal{ApJ}

\shorttitle{Protoplanetary to debris disks: separate evolution of the mm and $\mu$m-sized dust}
\shortauthors{Michel, van der Marel \& Matthews}

\begin{document}

\title{Bridging the gap between protoplanetary and debris disks: separate evolution of millimeter and micrometer-sized dust}

\correspondingauthor{Arnaud Michel}
\email{arnaud.michel@queensu.ca}

\author[0000-0003-4099-9026]{Arnaud Michel}
\affiliation{Department of Physics, Engineering Physics and Astronomy, Queen’s University, Kingston, ON, K7L 3N6, Canada}
\affiliation{Quest University Canada, 3200 University Boulevard, Squamish, BC, V8B 0N8, Canada}

\author[0000-0003-2458-9756]{Nienke van der Marel}
\affiliation{Department of Physics \& Astronomy, University of Victoria, Victoria, BC V8P 5C2}
\affiliation{Herzberg Astronomy \& Astrophysics Research Centre, National Research Council of Canada, 5071 West Saanich Road, Victoria, BC V9E 2E7, Canada}

\author[0000-0003-3017-9577]{Brenda C. Matthews}
\affiliation{Herzberg Astronomy \& Astrophysics Research Centre, National Research Council of Canada, 5071 West Saanich Road, Victoria, BC V9E 2E7, Canada}
\affiliation{Department of Physics \& Astronomy, University of Victoria, Victoria, BC V8P 5C2}

\begin{abstract}
 The connection between the nature of a protoplanetary disk and that of a debris disk is not well understood. Dust evolution, planet formation, and disk dissipation likely play a role in the processes involved. We aim to reconcile both manifestations of dusty circumstellar disks through a study of optically thin Class III disks and how they correlate to younger and older disks. In this work, we collect literature and ALMA archival millimeter fluxes for 85 disks (8\%) of all Class III disks across nearby star-forming regions. We derive millimeter-dust masses $M_{\text{dust}}$ and compare these with Class II and debris disk samples in the context of excess infrared luminosity, accretion rate, and age. The mean $M_{\text{dust}}$ of Class III disks is $0.29 \pm 0.19~M_{\oplus}$. We propose a new evolutionary scenario wherein radial drift is very efficient for non-structured disks during the Class II phase resulting in a rapid $M_{\text{dust}}$ decrease. In addition, we find possible evidence for long infrared protoplanetary disk timescales, ${\sim}8$~Myr, consistent with overall slow disk evolution. In structured disks, the presence of dust traps allows for the formation of planetesimal belts at large radii, such as those observed in debris disks. We propose therefore that the planetesimal belts in debris disks are the result of dust traps in structured disks, whereas protoplanetary disks without dust traps decrease in dust mass through radial drift and are therefore undetectable as debris disks after the gas dissipation. These results provide a hypothesis for a novel view of disk evolution.
\end{abstract}

\keywords{protoplanetary disks --- stars: pre-main sequence --- young stellar objects --- Kuiper belt: general --- submillimeter: general}

\section{Introduction} \label{sec:intro}

 The formation of planets around newly formed young stars is thought to be happening within the circumstellar disks of dust and gas, the protoplanetary disks. Studying the evolutionary properties and trends of such disks inform our greater understanding of the types of planets, planetary systems, and the remaining dust belts we observe today \citep{Armitage_2011,Mordasini_2012,Hughes_2018}. It also allows us to constrain dust dissipation and dust growth processes in the disks which inform planet formation theory and models \citep{Alibert_2005,Johansen_2007,Testi_2014}. The two key questions for this work are: `How do disks evolve?' and `Which mechanisms enable the protoplanetary disk's development into a debris disk?'.
 
 Observations of protoplanetary disks across multiple wavelengths from the infrared to the sub-mm/mm allow us to probe into the disk properties and help us develop an understanding of the evolutionary story of the disk \citep{Williams_2011,Wyatt_2015}. Observationally defined young stellar object (YSO) classification systems allow for the categorization and quantification of infrared excess emission emitted by the protoplanetary disk, such as the Lada classification \citep{Lada_1987,Greene_1994,Evans_2009}. A newly formed disk, still embedded in its parent envelope, is associated with a Lada Class 0/I YSO; as it evolves and the envelope disappears, the disk and star system becomes a gas-rich accreting primordial disk (Class II object), which further evolves into a gas-poor, non-accreting evolved disk (Class III object). At the latest stage, the gas and dust in the disk are almost completely dissipated. Depending on whether there is a significant reservoir of planetesimals and sufficient stirring, there can be a collisional cascade that leads to second-generation dust rings. Such systems are called debris disks \citep{Dunham_2015,Wyatt_2015}.
 
 These different classes can be linked to the theoretical understanding of the dust evolution processes taking place within the disks \citep{Evans_2009, Birnstiel_2010}. Many millimeter disk surveys have targeted protoplanetary disks in the Class 0 \& I phase \citep[e.g][]{Cox_2017,Segura-Cox_2018,Williams_2019,Tobin_2020,Tychoniec_2020}, the Class II phase \citep[e.g.][]{Andrews_2013,Ansdell_2016,Ansdell_2017,Barenfeld_2016,Cazzoletti_2019,Williams_2019}, but only a couple of these surveys have also targeted the evolved Class III disks \citep[e.g.][]{Hardy_2015,Barenfeld_2016,Williams_2019}. Recently, \citet{Lovell_2021} reported new ALMA observations of 30 Class III YSOs associated with the Lupus star forming region. Although they obtained very deep sensitivities for each target, only 4 Class III objects were detected. Based on their observational results, they suggest that there is rapid protoplanetary disk mm-dust dispersal and evolution such that planetesimals form within ${\sim}2$~Myr. Their conclusions suggest that some Lupus disks aged 1-3~Myr are already young debris disks and will evolve into objects similar to the currently observed debris disk sample (see Figure 10 in \citet{Lovell_2021}). Therefore, we aim to further expand on our understanding of the role of Class III disks in the evolutionary process bridging the gap between the protoplanetary and debris disk phases. To situate the disk history and context of Class III disks, this study also necessitates a reassessment of the Class II phase. 
 
 Previous studies show several important correlations in disk evolution during the protoplanetary disk phase: the mean millimeter dust mass of the disks decreases with the age of the star-forming region, with a stronger decrease for low-mass stars \citep{Ansdell_2017}; the dust disk radius decreases with age \citep{Hendler_2020}; the number of disks in a star-forming region drops with age, both according to infrared excess and number of accreting stars \citep{Hernandez_2007,Mamajek_2009,Fedele_2010,Ribas_2014}; accretion rates remain high even for older stars \citep{Rugel_2018,Venuti_2019,Manara_2020}, which are consistent with viscous evolution models when a low viscosity of $\alpha_{v}\footnote{We use the notation $\alpha_{v}$ to define the alpha viscosity, in contrast to the $\alpha_{Lada}$ and $\alpha_{IRAC}$ used for the infrared spectral slope elsewhere in this document.}=10^{-3}$ is assumed, in combination with dust evolution \citep{Sellek_2020}; millimeter dust masses of non-accreting young stars (WTTS) and debris disks are well below those of protoplanetary disks \citep{Wyatt_2008,Panic_2013,Hardy_2015}. Interestingly, several older star-forming regions contain one to a few bright, massive disks which lie well above the mean dust mass in that region \citep{Ansdell_2015,Ansdell_2020}, and, lastly, transition disks with large inner dust cavities have been suggested to be outliers in the general disk distribution \citep{Owen_2012,vandermarel_2018}, perhaps following a separate evolutionary path \citep{Currie_2009,vanderMarel_2021}.
 
 Disk evolution is intricately tied to dust growth, planet formation, and the dissipation of gas and dust. As the disks evolve through different stages, the various processes efficiencies vary and cumulatively either lock up dust and gas within planetesimals and planets, accrete material onto the host star, or dissipate these from the system. Within the disk, the effects of radial drift, viscous accretion onto the host star, photoevaporative and thermal winds, chemical evolution, and planet-disk interactions affect how the dust evolves and consequently, how the disks evolve too \citep[for recent reviews see][]{Alexander_2014,Testi_2014}. The development of sophisticated global dust evolution models \citep[e.g][]{Dullemond_2005,Tanaka_2005,Birnstiel_2010,Pinilla_2020,Sellek_2020} can simulate the evolution of dust grains and disks under a variety of parameters to explore how dust growth and removal processes are expected to impact the disks.
 
 In this work, we study Class III disks to explore the disk and dust evolution processes. Our sample is constructed from low-mass star-forming regions with well-defined membership in the solar neighborhood ($\lesssim300$~pc). In Section~\ref{sec:sample}, we present the star-forming region selection and explain the YSO classification methodology. In Section~\ref{sec:data}, we present millimeter dust masses, accretion rates, fractional disk luminosities, and star-forming region disk fractions using two different criteria (named the Lada and IRAC categorizations). In Section~\ref{sec:results}, we present the correlations between these parameters and the expected evolutionary trends connecting Class II, III, and debris disks. We also present updated Haisch-Lada plots (using the IRAC and Lada classification systems) exploring characteristic disk lifetimes based on infrared excesses. In Section~\ref{sec:discussion}, we present two different possible evolutionary pathways explaining how disks evolve and contextualize the work within our current understanding of disk evolution processes and most recent observations. Lastly, we provide a summary of our findings and open questions in Section~\ref{sec:summary}.
 
\section{Sample}\label{sec:sample}
\subsection{Target selection}
 We build a sample of Class III YSOs from low-mass nearby star-forming regions which have well-constrained membership studies, e.g. using \textit{Gaia} data. The selected regions span a range of ages from 1-12~Myr; these include Ophiuchus, Taurus, Chamaeleon I, Chamaeleon II, IC 348, Lupus (I, III, IV, V, \& VI), $\epsilon$ Chamaeleontis, Corona Australis, TW Hydra, $\eta$ Chamaeleontis, and Upper Sco. Regions such as Auriga, Cepheus, Musca, and Serpens are not included as they are either at much larger distances ($>$300 pc) and/or their stellar and disk information is not well constrained. Furthermore, we have purposefully not included any high UV-irradiated environments which are known to affect disk survival \citep[e.g.][]{Haworth_2017}. This selection is intentional to build a sample with comparatively similar external influences.

 Class III YSOs have been known to be confused with background contaminants, e.g. galaxies or AGB stars, with similar infrared excesses as they demonstrate similar expected signatures \citet{Allen_2004,Harvey_2007,Oliveira_2009,Evans_2009}. However, we are now able to disentangle these much more easily thanks to the advent of the second data release of the \textit{Gaia} mission \citep{GaiaDR2_2018}. \citet{Manara_2018} made an excellent demonstration with Lupus V \& VI using \textit{Gaia} parallax, and thus distances, to confirm and reject previously identified YSOs as to their membership to the star-forming regions. More recently, complete \textit{Gaia}-selected membership studies are redefining star-forming regions in great detail using high precision astrometry and photometry \citep[e.g.][and references therein]{Herczeg_2019,Luhman_2020_lupus,Galli_2020_cra}. As such, we outline briefly in the Appendix how each of the eleven star-forming regions' YSOs have been selected to confirm their membership to their specific region. We also specify the source of the infrared photometry that we use. The infrared data are preferentially, when available, chosen from \textit{Spitzer} IRAC/MIPS1 bands due to the higher sensitivity
 and if lacking then these are from \textit{WISE}~1-4 bands \citep{Avenhaus_2012}. The complete list and details per region can be found in Appendix~\ref{appendix:regions}.

\subsection{Classification}\label{sec:classification}

 Class III disks are identified using the infrared spectral index $\alpha_{\rm Lada}$ between 2 and 22-24 $\mu$m ($K$-band and \textit{Spitzer}-MIPS1 or WISE4), following \citet{Greene_1994}:
\begin{itemize}[noitemsep]
    \item Class I: $\alpha_{\rm Lada}>0.3$
    \item Class F: -0.3 $<\alpha_{\rm Lada}<$ 0.3
    \item Class II: -1.6 $<\alpha_{\rm Lada}<$ -0.3
    \item Class III: $\alpha_{\rm Lada}<$ -1.6
\end{itemize}
 where the empirical SED classes generally correspond to physical evolutionary stages \citep{Evans_2009,Dunham_2014}. The $\alpha_{\rm Lada}$ is computed from $K$-MIPS1/WISE4 where possible classifying the YSOs as Class I, F, II, or III. However, a series of YSOs, while detected at shorter infrared wavelengths (2 to 12~$\mu$m - IRAC1/WISE1 to IRAC4/WISE3 bands) are lacking 22-24~$\mu$m (MIPS1/WISE4) detections, given that these are members of young star-forming regions, we define these as Class III-short$\lambda$ (Class III-s$\lambda$). We avoid using the term diskless or bare photospheres as recent studies \citep{Lovell_2021} demonstrate that evolved disks can have very faint and low mass disks.
 
 Secondly, all targets were classified using an alternative classification using only the IRAC/WISE1-3 bands. We define this as IRAC classification, $\alpha_{\rm IRAC}$, as it is solely based on the infrared slope of the IRAC/WISE1-3 wavelengths \citep{Lada_2006}. This parameter has been used to estimate the typical disk lifetime \citep{Lada_2006,Hernandez_2007,Mamajek_2009}. The classification has been adapted for the lastest/most evolved stage, it is as follows:
\begin{itemize}[noitemsep]
    \item $\alpha_{\rm IRAC}>0$: protostar candidates 
    \item $-1.8<\alpha_{\rm IRAC}<0$: disk-bearing stars
    \item $-2.56<\alpha_{\rm IRAC}<-1.8$: anemic disk-bearing stars
    \item $\alpha_{\rm IRAC}<-2.56$: near diskless stars.
\end{itemize}

 For targets with incomplete photometry, $\alpha_{\rm IRAC}$ is computed as well as possible from the available bands with priority being given to a computation from IRAC bands rather than WISE1-3 due to the better sensitivity of \textit{Spitzer} observations. Table \ref{tbl:regiontable} presents the final number of targets in each region. In total, 976 Class III objects are identified across 3741 young stars in 11 star-forming regions. 

\begin{table*}[!ht]
\begin{center}
\caption{Star-forming regions in the sample}
\label{tbl:regiontable}
\begin{tabular}{llll|llllll|ll}
\hline
&&&& & \multicolumn{2}{|c|}{Class III} & Class II & Cl I+F & Class III-s$\lambda$ & \multicolumn{2}{c}{P\textsubscript{diskfrac}} \\
Region & Age & $d$ & Ref. & $N_{\rm total}$ & $N$& $N_{mm}$ & $N$ & $N$ & $N$ & IR & Lada \\
&(Myr)&(pc)&(0)&(1)&(2)&(3)&(4)&(5)&(6)&(7)&(8)\\
\hline
\hline
Ophiuchus & 1-2 & 139 & 1 & 420 & 134 & 13 & 168 & 92 & 26 & $56 \pm 5$ & $62 \pm 5$ \\
Taurus & 1-2 & 128-198 & 2 & 467 & 91 & 12 & 167 & 62 & 147 & $46 \pm 4$ & $49 \pm 4$\\
Cham I & 1.7-2.4 & 187-194 & 3 & 183 & 81 & 11 & 81 & 21 & - & $76 \pm 18$ & $76 \pm 18$ \\
Cham II & 1.7-2.3 & 198 & 4 & 41 & 10 & 0 & 24 & 7 & - & $51 \pm 7$ & $56 \pm 7$ \\
IC 348 & 2-3 & 310 & 5 & 349 & 17 & 1 & 112 & 26 & 194 & $50 \pm 5$ & $40 \pm 4$ \\
Lupus & 2.6-3.1 & 160 & 6 & 196 & 20 & 8 & 85 & 19 & 72 & $51\pm 6$ & $ 53\pm 6$ \\
$\epsilon$ Cha & 3-8 & 110 & 7 & 40 & 25 & 0 & 10 & 2 & 3 & $23 \pm 9$ & $30 \pm 10$ \\
CrA & 5-6 & 160 & 8 & 275 & 184 & 2 & 77 & 14 & - & $18 \pm 3$ & $28 \pm 3$ \\
TW Hya & 7-13 & 56 & 9 & 40 & 26 & 2 & 11 & 1 & 2 & $25 \pm 9$ & $30 \pm 10$\\
$\eta$ Cha & 8-14 & 94 & 10 & 18 & 12 & 2 & 5 & 1 & - & $28 \pm 14$ & $33 \pm 16$\\
Upper Sco & 10.5-12 & 145 & 11 & 1712 & 376 & 34 & 291 & 8 & 1037 & $17 \pm 1$ & $17 \pm 1$\\
\hline
Total & & & & 3741 & 976 & 85 & 1031 & 253 & 1481 & & \\
\hline
\end{tabular}
\\
\end{center}
(0) References for the age and distance of the star-forming regions: 1) \citet{Wilking_2008,Esplin_2020}; 2) \citet{Kraus_2009,Galli_2019}; 3) \citet{Galli_2021} 4) \citet{Galli_2021}; 5) \citet{Luhman_2003,Ruiz-Rodriguez_2018} 6) \citet{Galli_2020_lupus} 7) \citet{Murphy_2013} 8) \citet{Galli_2020_cra} 9) \citet{Bell_2015,Weinberger_2013} 10) \citet{Bell_2015,Lyo_2004} 11) \citet{Esplin_2018,Luhman_2020_usco} 
\\
(1) Total number of YSOs in their respective star-forming regions including Class I+F, II, III, and III-short$\lambda$ objects.\\
(2) Number of Class III objects. \\
(3) Number of Class III objects with available millimeter fluxes from ALMA.\\
(4) Number of Class II objects. \\
(5) Number of Class I and Flat objects. \\
(6) Number of Class III-short$\lambda$ objects. That is YSOs which have been at least partially detected in IRAC or WISE bands but have no MIPS1/WISE-4 detections. \\
(7) Frequency of thick disk-bearing YSOs according to the criterion -1.8 $<\alpha_{\rm IRAC}$ (see text). \\
(8) Frequency of protoplanetary disks with infrared excess qualifying these as Class I, Flat, and II disks. \\
\end{table*}

\section{Data} \label{sec:data}

 To study Class III disks ALMA millimeter fluxes and stellar information are collected. Spectral types are also collected from the literature (see Table~\ref{tbl:mastertable} for references) and any stars earlier than A0 or later M6 are removed for a proper comparison with previous protoplanetary disk studies.

\subsection{ALMA millimeter flux}\label{sec:alma}

 We collected ALMA 890~$\mu$m and 1.3~mm continuum fluxes and upper limits from the literature to derive the mm-dust disk masses, $M_{\text{dust}}$ of Class III disks. References for literature values are provided in Table~\ref{tbl:mastertable}. 14 objects across our star-forming regions have been observed with ALMA, but the data are unpublished, these are presented in Table~\ref{tbl:alma_archival}. We reduce these ALMA archival data-sets using the provided \texttt{CASA} reduction scripts and image the targets using the \texttt{CLEAN} algorithm with natural weighting \citep{McMullin_2007}. Three disks are detected and we measure the total flux using the \texttt{uvmodelfit} task; for the remaining 11 non-detections, 3$\sigma$ upper limits are derived. In total, from the literature and unpublished archival observations, we collect 85 ALMA fluxes (including 64 upper limits) out of the sample of 976 Class III objects. 

\begin{table*}
    \caption{ALMA archival data of unpublished Class III disks}  
    \label{tbl:alma_archival}
    \centering
    \begin{tabular}{l c c c c c c c c}
    \hline
    2MASS & ALMA program & PI & Band & Beam & RMS & Bandwidth & Frequency & $F_{mm}$ \\
     &  &  &  & ('') & (mJy) & (GHz) & (GHz) & (mJy) \\
    \hline
    J16082843-3905324 & 2011.0.00733.S & M. Schreiber & 6 & $0.57\times0.57$ & 0.16 & 3.75 & 232.4 & $<0.48$ \\
    J16083156-3847292 & 2011.0.00733.S & " & 6 & $0.55\times0.55$ & 0.13 & 3.75 & 232.4 & $1.00 \pm 0.13$ \\
    J11045100-7625240 & 2012.1.00313.S & L. Testi & 7 & $0.55\times0.38$ & 0.13 & 7.5 & 342.2 & $<0.38$ \\
    J11124299-7637049 & 2012.1.00313.S & " & 7 & $0.55\times0.38$ & 0.13 & 7.5 & 342.2 & $<0.39$ \\
    J11091172-7729124 & 2013.1.01075.S & S. Daemgen & 7 & $0.75\times0.39$ & 0.35 & 6 & 339.3 & $<1.04$ \\
    J11145031-7733390 & 2013.1.01075.S & " & 7 &  $0.73\times0.39$ & 0.32 & 6 & 339.3 & $<0.95$ \\
    J11075588-7727257 & 2013.1.01075.S & " & 7 &  $0.74\times0.39$ & 0.35 & 6 & 339.3 & $<1.05$ \\
    J04332621+2245293 & 2016.1.01511.S & J. Patience & 7 & $0.14\times0.13$ & 0.12 & 6 & 336.5 & $<0.36$ \\
    J04354203+2252226 & 2016.1.01511.S & " & 7 & $0.14\times0.13$ & 0.06 & 6 & 336.5 & $0.49 \pm 0.06$ \\
    J04331003+2433433 & 2017.1.01729.S & W. Wang & 6 & $0.31\times0.18$ & 0.013 & 7.5 & 225.0 & $<0.038$ \\
    J11062877-7737331 & 2017.1.01627.S & C. Caceres & 6 & $0.52\times0.32$ & 0.01 & 6.3 & 282.8 & $1.33 \pm 0.01$ \\
    J16130627-2606107 & 2018.1.00564.S & J. Carpenter & 7 & $0.31\times0.31$ & 0.14 & 7.5 & 334.2 & $<0.42$ \\ 
    J16114612-1907429 & 2018.1.00564.S & " & 7 & $0.31\times0.31$ & 0.14 & 7.5 & 334.2 & $<0.53$ \\
    J16191936-2329192 & 2018.1.00564.S & " & 7 & $0.32\times0.32$ & 0.14 & 7.5 & 334.2 & $<0.53$ \\
    \hline
    \end{tabular}
\end{table*} 

 The ALMA sub-mm/mm fluxes are used to calculate the mm-dust disk mass, $M_{\text{dust}}$, using Eq.~(1) from \citet{Ansdell_2016} for 890~$\mu$m fluxes and Eq.~(1) from \citet{Ansdell_2018} for 1.3~mm fluxes. These equations are based on the assumption that the dust emission at sub-mm/mm wavelengths is isothermal and optically thin as demonstrated by \citet{Hildebrand_1983}:
\begin{equation}
    M_{\text{dust }}=\frac{F_{\nu} d^{2}}{\kappa_{\nu} B_{\nu}\left(T_{\text{dust}}\right)}
\end{equation}
 where $B_{\nu}$ is the Planck function for a characteristic dust temperature, $T_{\text{dust}}$, the dust grain opacity, $\kappa_{\nu}$, the distance to the target in parsecs, $d$, and the sub-mm/mm flux, $F_{\nu}$. For consistency with protoplanetary disk studies and statistics, the same parameters are used as in \citet{Ansdell_2016,Ansdell_2018}; $T_{\mathrm{dust}} = 20~\mathrm{K}$, $k_{v} = 10~\mathrm{cm}^2\mathrm{g}^{-1}$ at 1000~GHz, and an opacity power-law index of $\beta = 1$. For the distance, $d$, we used the inverse of the \textit{Gaia} parallax. For targets without reported \textit{Gaia} parallax, the average distance to the star-forming region is assumed, see Table~\ref{tbl:regiontable}. The resulting mm-dust disk masses can be found in Table~\ref{tbl:mastertable}. 

\subsection{Fractional Disk Luminosity}\label{sec:fracd}
 A second relevant parameter in our study is the fractional disk luminosity, $L_{\text{fract}}$ = $L_d/L_*$, derived from the SED from optical to infrared wavelengths. We define the disk luminosity $L_d$ as the difference in flux between de-reddened observations and the stellar photosphere model, integrated over the range of 1.66 to 110~$\mu$m, see Appendix~\ref{appendix:Lfract_SED} for examples including a Class II, a Class III and a Class III transitional disk (Sz~91).
 We also demonstrate there that the fractional luminosity is generally dominated by the excess at shorter wavelengths.

 SEDs are constructed using data from \textit{Spitzer} IRAC/MIPS, WISE~1-4, 2MASS \textit{JHK}, and optical \textit{GBVI} photometry, where available. Specific references for the Spitzer photometry are provided in Section \ref{sec:sample}. The data are de-reddened using the \citet{Cardelli_1989, O'Donnell_1994} extinction law assuming $R_V=3.1$ and scaled to the visual extinction, $A_V$. Spectral types, stellar luminosities $L_*$, and visual extinction are collected from the literature and scaled to the new \textit{Gaia} distances. For some targets, the stellar luminosity and extinction are re-fit to ensure the best fit between the model photosphere and photometric data. For the targets that were re-fit, we aim to ensure that the luminosities and extinction used are within uncertainties reported in the literature, when that is not the case the luminosity and/or extinction is flagged. The stellar properties and flags are provided in Table~\ref{tbl:mastertable}. Using these properties, stellar photosphere models are constructed using Kurucz models for spectral types between A0 and K7 \citep{Castelli_2003}, and BT-Settl models for M0 to M6 stars \citep{Allard_2012, Allard_2013}. The fractional luminosity is derived accordingly (listed in Table~\ref{tbl:mastertable}) and the SEDs for the Class III targets used are provided in Figure \ref{fig:seds} in Appendix~\ref{appendix:SEDs}.

\subsection{Accretion rates}
\label{sec:accretionrates}
 A third relevant parameter for disk evolution is the stellar accretion rate, $\dot{M}_{\text{acc}}$. Young stars are generally classified as either Classical T Tauri stars (CTTS) or Weak-line T Tauri stars (WTTS), depending on their accretion rate (typically a threshold of ${\sim}10^{-11}~M_{\odot} yr^{-1}$,  following \citet{Natta_2004}). This distinction is based on the width of the H$\alpha$ line, both the equivalent width and the width at 10\% of the maximum \citep{WhiteBasri2003}, and in recent works the UV excess is also considered \citep[e.g.][]{Ingleby2011,Manara_2013_xshooter,Thanathibodee2018}. Although both WTTS and Class III objects are considered evolved protoplanetary disks, they are not the same. \citet{Wahhaj_2010} found that most WTTS have a [K]-[24] color of 0$ \pm $0.15, where CTTS have [K]-[24]$\geq$2. The WTTS color corresponds to an $\alpha_{\text{Lada}}$=-2.8, well below the threshold of Class III disks ($<$-1.6). This means that WTTS essentially cover the lowest infrared range of Class III objects and these two populations cannot be compared directly. Therefore, some Class III objects may still be accretors. In this section, we aim to compare the trends of accretion rates of Class II objects with that of Class III objects, for the targets where accretion rates are  available. 
 
 Inspection of the literature for our Class III objects confirms that 80\% of the targets in the younger regions are WTTS. YSOs in Upper Sco have not been studied systematically for H$\alpha$ emission and their accretion status thus remains unknown, although it is very likely that the Class III objects are primarily WTTS as well.
 
 However, the H$\alpha$ width only provides a reliable estimate of accretion status: accretion rates based on emission lines are generally much more uncertain, due to contributions from winds, rotation, and chromospheric activity \citep{Manara_2013_xshooter,Hartmann_2016}. Therefore, in this work we only consider accretion rates derived using broadband flux-calibrated spectra from UV-NIR, which takes into account both the UV excess and line contributions to measure the accretion luminosity $L_{\rm acc}$ \citep[e.g.][]{Herczeg_2008, Alcala_2014} so that all accretion rates are derived in the same manner.

This limits the regions with known accretion rates constrained using \textit{Gaia} distances to:
\begin{itemize}[noitemsep]
    \item Lupus \citep[][55 CII, 3 CIII]{Alcala_2017}
    \item Chamaeleon I \citep[][4 CIII]{Manara_2017}
    \item $\eta$ Cha \citep[][6 CII, 9 CIII]{Rugel_2018}
    \item TW Hya \citep[][3 CII, 2 CIII]{Venuti_2019} 
    \item Upper Sco \citep[][30 CII, 2 CIII]{Manara_2020}.
\end{itemize}

 For Lupus and Chamaeleon I, the stellar parameters have been updated with the new \textit{Gaia} distances \citep{Manara_2018}. We collect these to conduct a comparison of the $\dot{M}_{\text{acc}}$ of Class II disks between one younger star-forming region (Lupus) and older ones (mainly Upper Sco but also including $\eta$ Cha, and TW Hya). 
 Therefore, Class II disks in Chamaeleon I from \citet{Manara_2017} are not included in our comparison as their stellar properties and trends ($\dot{M}_{\text{dust}}$ - $M_{*}$ and $\dot{M}_{\text{acc}}$ - $M_{*}$) are very similar to those of Lupus \citep{Pascucci_2016,Manara_2020}. 
 The additional Class III targets with $\dot{M}_{\text{acc}}$ values are collected for reference but are not used in the analysis due to the significant incompleteness of the samples.

\subsection{Disk fraction}
\label{sec:diskfraction}
 The infrared fluxes of YSOs are used to derive an estimate of the disk fraction in each star-forming region, using both the $\alpha_{\rm Lada}$ spectral index \citep{Greene_1994} between 2.2 and 24 $\mu$m (Lada classification, defined above) and the $\alpha_{\rm IRAC}$ spectral index between 3.6 and 8.0 $\mu$m (IRAC classification). As the star-forming regions' memberships have been re-evaluated with \textit{Gaia}, the Class III and Class III-short$\lambda$ samples have been amended and the disk fractions have changed compared to previous measurements. The disk fraction is defined by the IRAC classification as the number of protostars and disk-bearing stars divided by the total number of YSOs. In the Lada classification, we define the disk fraction as the number of Class I+F and II disks divided by the total number of YSOs. Both disk fractions are computed per star-forming region and provided in Table \ref{tbl:regiontable}. 

\section{Results}\label{sec:results}

\begin{figure}[!ht]
   \centering
   \includegraphics[]{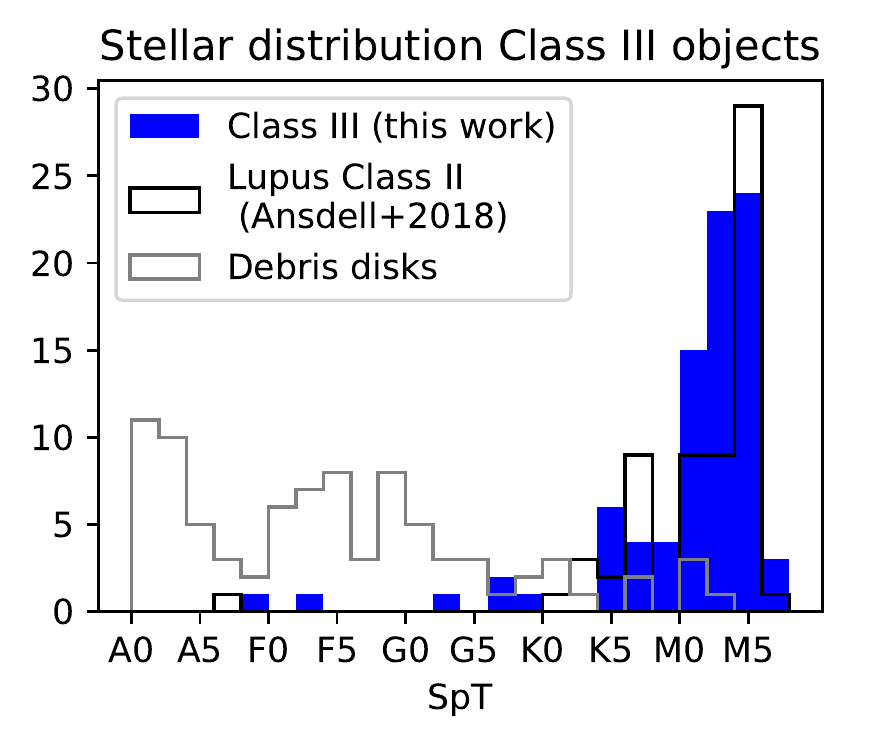}
      \caption{Distribution of spectral types of the Class III objects with ALMA observations, compared with the spectral type distribution of the Class II survey of the Lupus star forming region \citep{Ansdell_2018} and of the sample of debris disks in the SONS survey \citep{Holland_2017} as described in the text. This comparison shows that our Class III sample is not biased towards early spectral types and is comparable to a typical cluster distribution, whereas the debris disk sample is biased towards early spectral types.}
         \label{fig:sptdist}
 \end{figure}

 Dust masses are calculated for 85 Class III objects. Most of these (64) are upper limits, but the upper limits vary broadly due to the different sensitivities of the different ALMA programs. The deepest limit is ${\sim}10^{-2}~\text{M}_{\oplus}$, whereas the highest detection is 1.5~$\text{M}_{\oplus}$ (excluding Sz~91 which is a known transition disk). The ALMA sample only contains a fraction (${\sim}8\%$) of all Class III objects in the nearby star-forming regions and is non-uniform due to the random availability of ALMA fluxes. Nevertheless, this sample represents a wide range of spectral types and stellar masses comparable to the typical spectral type distribution in Lupus (Figure \ref{fig:sptdist}). This comparison shows that our Class III sample, although it consists of a random selection of Class III targets from the ALMA archive, is not biased towards early spectral types and is comparable to a typical cluster distribution, unlike the debris disks from the SONS survey as described below. Given that we find detections in most regions (Table \ref{tbl:regiontable}, column $N_{mm}$) we consider the Class III sample to be representative for comparisons with less and more evolved disk populations.
 
\begin{figure*}[!ht]
    \gridline{\fig{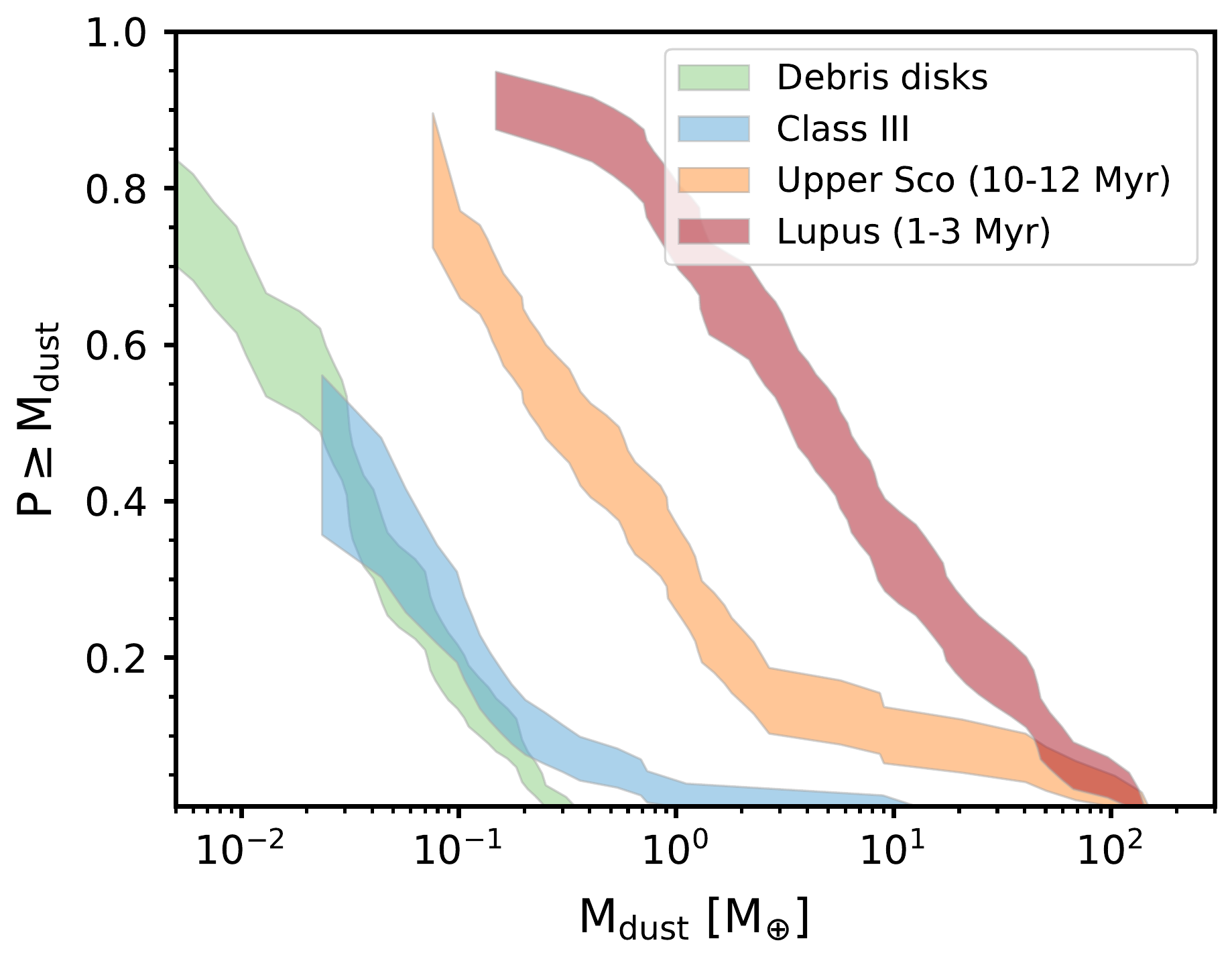}{0.5\textwidth}{(a)}
          \fig{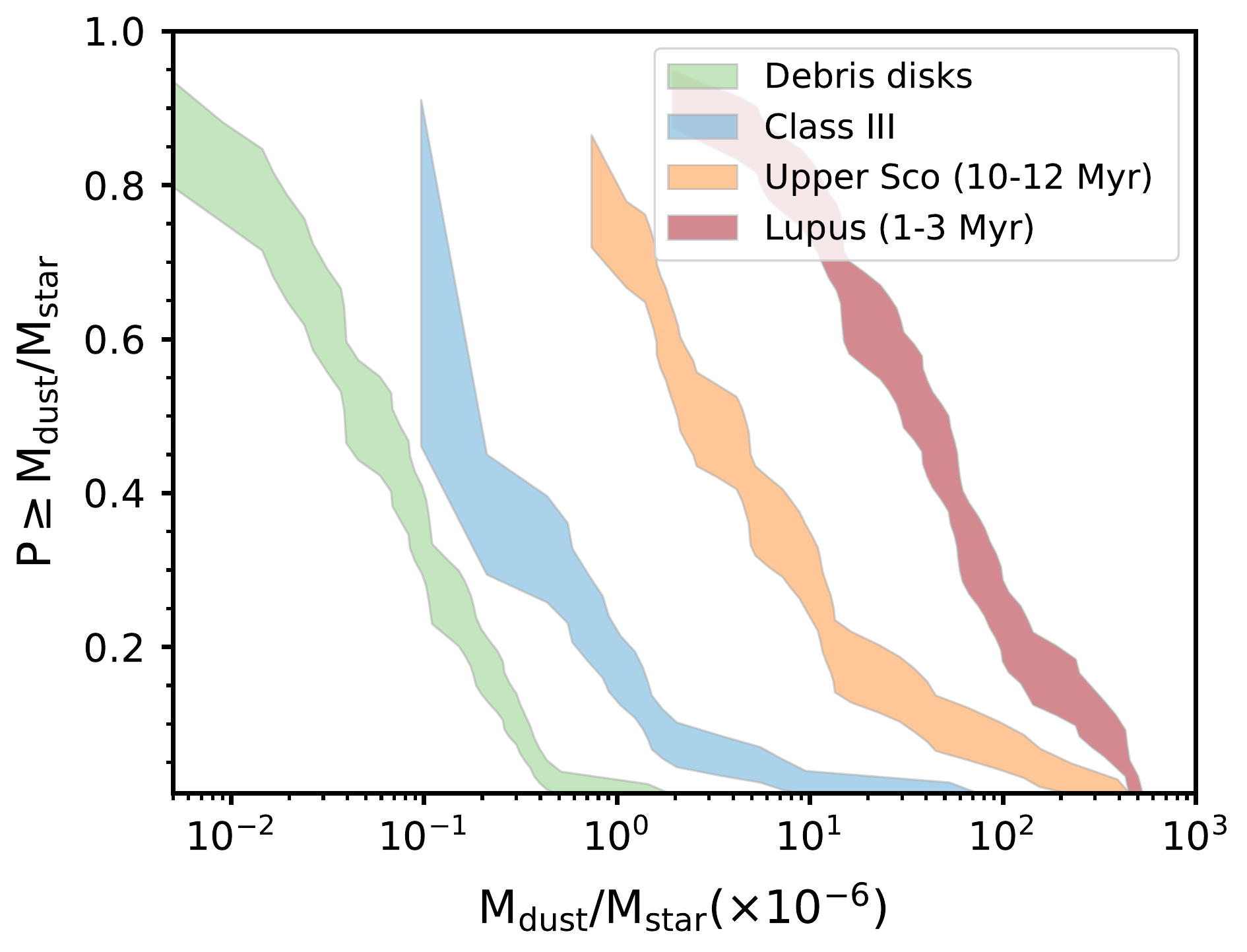}{0.5\textwidth}{(b)}
          }
    \caption{{\bf Left:} The dust mass cumulative distribution for the Class II disks in Lupus (in red) \citep{Ansdell_2016} and Upper Sco (in yellow) \citep{Barenfeld_2016}, all Class III disks (in blue), and the debris disk sample from the SONS survey (in green) \citep{Holland_2017}. {\bf Right:} The disk-to-star ($M_{\text{dust}}/M_*$) cumulative distributions using the same colours as the left panel, in order to take into account the different stellar mass distribution in the SONS sample. For both panels, the distributions are constructed using the Kaplan-Meier estimator, it includes the upper limits and the line widths are based on a $1\sigma$ confidence. \label{fig:asurv}}
 \end{figure*}

 Figure~\ref{fig:asurv}a shows the cumulative distribution of the millimeter dust mass ($M_{\text{dust}}$) for the Class III disk population calculated using the Kaplan-Meier estimator (KME) from the ASURV package to include upper limits \citep{Lavalley_1992}. To situate the Class III disks within their dust evolutionary context we have added 63 Class II objects from Lupus \citep{Ansdell_2016} and 69 from Upper Sco  \citep{Barenfeld_2016,Andrews_2018,Williams_2019} that are validated as members by the Lupus \citep{Luhman_2020_lupus} and Upper Sco \citep{Luhman_2020_usco} \textit{Gaia} membership studies. We note that the Upper Sco disk sample by \citet{Barenfeld_2016} was selected using a different disk criterion \citep{Luhman_2012} than the $\alpha_{\rm Lada}$ Class II criteria, and actually contains 67 Class II objects and 28 Class III objects, plus 9 objects without full infrared photometry. We have split these according to their Lada classification in our sample. Furthermore, the Class II objects in \citet{Barenfeld_2016} were limited to spectral types G2-M5. We have added to the \citet{Barenfeld_2016} Upper Sco Class II sample 2 targets from \citet{Andrews_2018} as well as 6 targets from \citet{Williams_2019} but identified as Upper Sco Class II targets by \citet{Luhman_2020_usco} (see Table \ref{tbl:mastertable_classii}). 
 
 For comparison, we add a large cold debris disk sample from sub-millimeter observations \citep{Holland_2017}. Figure~\ref{fig:sptdist} illustrates how the debris disk sample is biased toward early type stars compared to the Class II and III samples. The source of this bias lies in the relative detectability of debris disks and, likely, the higher occurrence of debris disks around early type stars \citep{Sibthorpe_2018}, both of which result in enhanced detection rates for debris disks around early type stars. As a result, in contrast to the Class II samples, the debris disk sample is not a complete population, and it is not representative of the initial mass function (IMF) that is skewed toward late type stars. The SONS survey sample targeted known debris disks of sufficient brightness to be detectable with SCUBA-2 at the JCMT, based on existing detections in the mid-IR and far-IR, favoring disks around earlier type stars. Therefore, our sample of the outcomes of disk evolution processes is biased very differently than the sample of YSOs available in nearby star-forming regions. The debris disk sample we use is based on the SONS survey which is composed of 84 disks: 46 detections and 38 upper limits. While \citet{Holland_2017} presented SED fitting results for SONS detections, here we use SED fits to all observed targets, including non-detections (G. Kennedy, private comm.). The input data and SED fitting method is essentially the same as in \citet{Holland_2017}, but uses updated software as outlined in \citet{Yelverton_2019}. Note that a small number of targets observed by SONS are not included here, as it was later concluded that their IR excesses were spurious (e.g. based on new Herschel observations), these include: HD 59601, HD~91312, HD~91782, HD~135502, HD~139590, HD~149630. We also exclude HD~98800 as it has been identified as a quadruple star system in which HD~98800B hosts a disk with a high $L_{\text{fract}}{\sim}10^{-1}$ and a CO detection, such that both the gas and dust are likely optically thick, qualifying it as a protoplanetary disk \citep{Koerner_2000,Holland_2017,Kennedy_2019}. Lastly, for the consistency of the Class II and III disk comparisons with the debris disk sample, we exclude B-type stars from our analysis.

 Figure \ref{fig:asurv}b shows the dust mass normalized by the stellar mass ($M_{\text{dust}}/M_*$ ratio) to illustrate the dust mass distribution with respect to the stellar host mass. This provides a better comparison with the debris disks as this sample is biased towards early-type massive stars and their stellar mass distribution is thus skewed with respect to the other disk distributions.

 According to the KME calculation, the mean dust mass of Class II disks is $18.7 \pm 4.0~\text{M}_{\oplus}$ in Lupus and $8.1 \pm 3.2~\text{M}_{\oplus}$ in Upper Sco, for all Class III disks we calculate $0.29 \pm 0.19~\text{M}_{\oplus}$, and for debris disks, it is $0.053 \pm 0.008~\text{M}_{\oplus}$. The mean $M_{\text{dust}}/M_* \times 10^{-6}$ found are $98.4 \pm 17.6$, $24.8 \pm 8.7$, $1.8 \pm 1.1$, and $ 0.126 \pm 0.028$ for Lupus Class II, Upper Sco Class II, all Class III, and debris disks, respectively. For the two panels in Figure~\ref{fig:asurv}, we have decided to group all Class III targets and all debris disk targets; we do not separate them by region or average age, as both the Class III and debris disk samples are not intentionally uniformly constructed and contain many upper limits.

\subsection{Dust mass evolution with \texorpdfstring{$L_{\mathrm{fract}}$}{}}

\begin{figure*}[!ht]
   \centering
   \includegraphics[width=\textwidth]{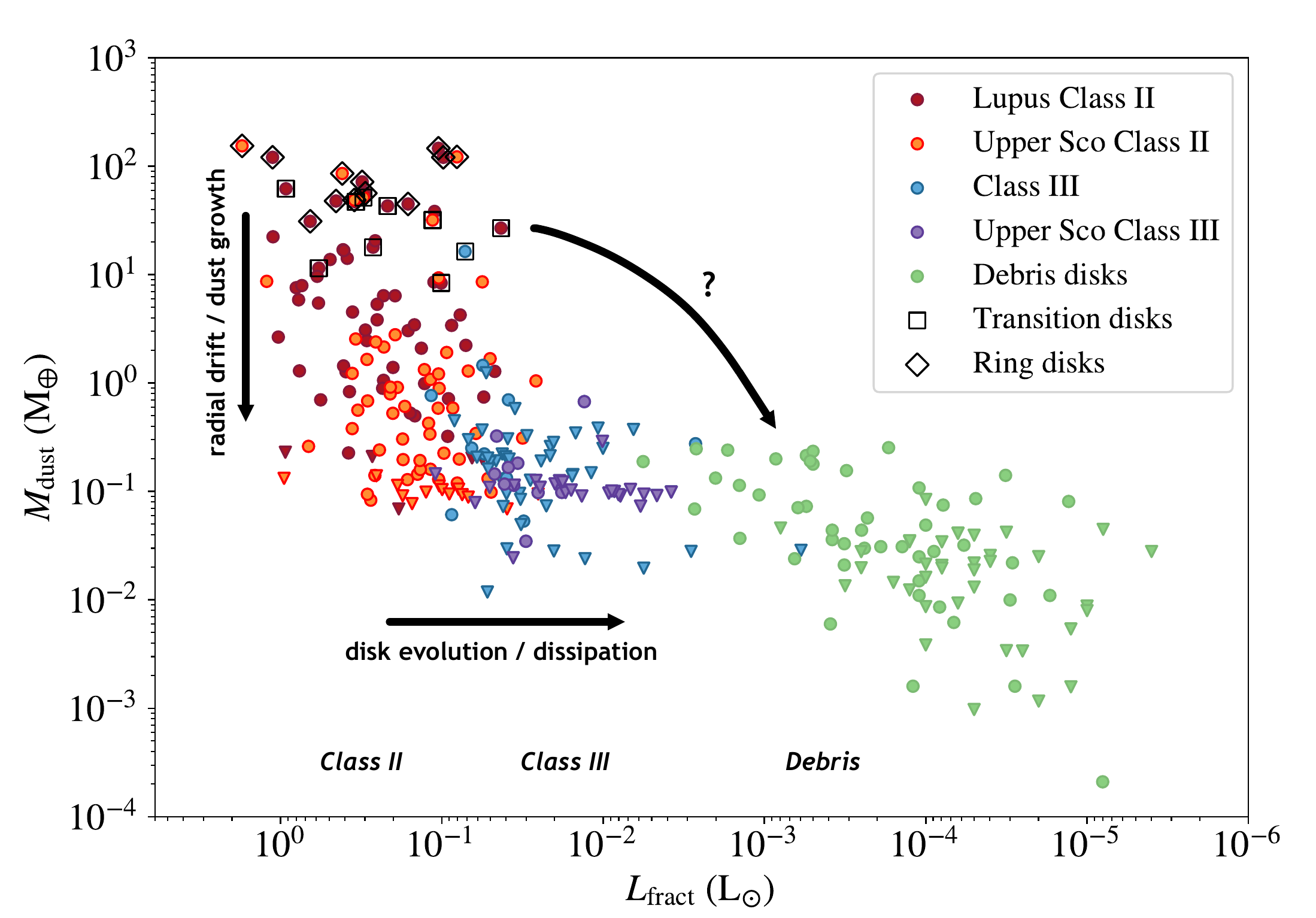}
      \caption{Fractional disk luminosity, $L_{\text{fract}}$, vs mm-dust disk mass, $M_{\text{dust}}$, for different evolutionary stages across star-forming regions. The same colour scheme as Figure~\ref{fig:asurv} is applied with the exception that Class III disks are separated into two bins between Upper Sco (in purple) and all other regions (in blue). The round scatter points are detections in the mm-continuum observations while downward facing triangles are non-detections; they are calculated based on a 3$\sigma$ flux. The structured ring and transition disks are identified by squares and diamonds and the mean structured $M_{\text{dust}}$ is 62~M$_{\oplus}$. The black arrows indicate the disk evolution mechanisms driving the overall evolutionary trends in protoplanetary disks. As dust growths and drifts inwards, $M_{\text{dust}}$ decreases and as disk dissipation takes place, $L_{\text{fract}}$ declines. The third arrow at the top connects the structured disks with the debris disks indicating a possible evolutionary pathway. The different disks' evolutionary stages, Class II, Class III, and debris disks, are noted to highlight how these regimes correlate with multiple observationally-defined features as explained in the text.}
         \label{fig:lfractmdust}
\end{figure*}

 To investigate the disk dust evolution process, we plot in Figure \ref{fig:lfractmdust} the millimeter dust mass as a function of the fractional disk luminosity, $L_{\text{fract}}$, for our Class III sample and the other disk distributions. The fractional disk luminosity traces the infrared emission of micron-sized dust of the disk \citep{Kenyon_1987,Hughes_2018} while the millimeter observations enable us to study the mass of the millimeter grains.
 
 Ages of individual young stars are highly uncertain for $<$10 Myr, due to extinction and differences between stellar evolutionary models \citep{Bell_2015}, so we have chosen not to plot the targets as a function of age unlike previous works \citep{Wyatt_2008, Hardy_2015, Wyatt_2015}. The uncertainties and inadequacies of the age dependencies are further discussed in Section \ref{subsec:ages}. The fractional luminosity values are typically on the order of ${\sim}10^{-1}$ for protoplanetary disks \citep{Cieza_2010} and $\leq10^{-3}$ for debris disks \citep{Wyatt_2008} and the only reliable distinction between these two types of disks \citep{Matthews_2014} other than an unambiguous age of the central star. Although the infrared emission is likely marginally optically thick in the protoplanetary disk phase and the origin of the dust is likely different for these two disk phases, it provides a way to compare the different stages in the dust evolution process. 

 Figure \ref{fig:lfractmdust} shows a clear decrease of millimeter dust mass $M_{\text{dust}}$ beyond $L_{\text{fract}}\leq10^{-1}$, the approximate borderline between Class II and III disks. In the high fractional luminosity regime ($L_{\text{fract}}{\sim}10^0$ to $10^{-1}$, the Class II disk phase) there is a large distribution of mm-dust masses over at least 3 orders of magnitude, where 95\% of $M_{\text{dust}}$ fall between $0.1$ and $120~\text{M}_{\oplus}$. On the other hand, the disks with $L_{\text{fract}}$ within the $10^{-1}$ to $10^{-3}$ regime (Class III) have a dust mass that goes up to 1.5 $\text{M}_{\oplus}$ (excluding Sz~91, see Section \ref{sec:structured}), and 95\% of observed $M_{\text{dust}}$ values are between $0.02$ and $1.2~\text{M}_{\oplus}$, overlapping with the lower end of the Class II disk dust masses. Due to the incompleteness and the large fraction of upper limits in the Class III sample, the full range of dust masses may be several orders of magnitude as well into the lower dust masses. The upper range of dust masses for the debris disk is comparable to the Class III objects; the closer proximity of the majority of the debris disks means the lower range of detected dust masses extends to 0.001 $\text{M}_{\oplus}$.

 Although there is a decline in $M_{\text{dust}}$ between Class II and Class III disks (higher and lower $L_{\text{fract}}$, respectively), a global correlation between $M_{\text{dust}}$ and $L_{\text{fract}}$ is not possible to assess due to the incompleteness and sensitivity limits of the more evolved Class III sample in comparison to the Class II samples. However, different regimes can be identified, possibly related to different evolutionary processes. Several aspects of our work focus on the analysis of the Class II objects for a proper understanding of the transition to Class III and debris disks.

\paragraph{Class II regime} 
 Within the high fractional luminosity regime, the mean millimeter dust masses of the disks in the older Upper Sco region are lower than those of the younger Lupus disks \citep{Barenfeld_2016,Ansdell_2017}. The KME computed mean Class II Lupus disk mass is $18.7 \pm 4.0~\text{M}_{\oplus}$ while Class II Upper Sco disks are around $8.1 \pm 3.2~\text{M}_{\oplus}$. However, there is no significant difference in the fractional luminosity values between these two regions. The Class II mean $L_{\text{fract}}$ is $0.33\pm0.27$ and $0.24\pm0.29$ for Lupus and Upper Sco, respectively.

\begin{figure*}[!ht]
 \gridline{\fig{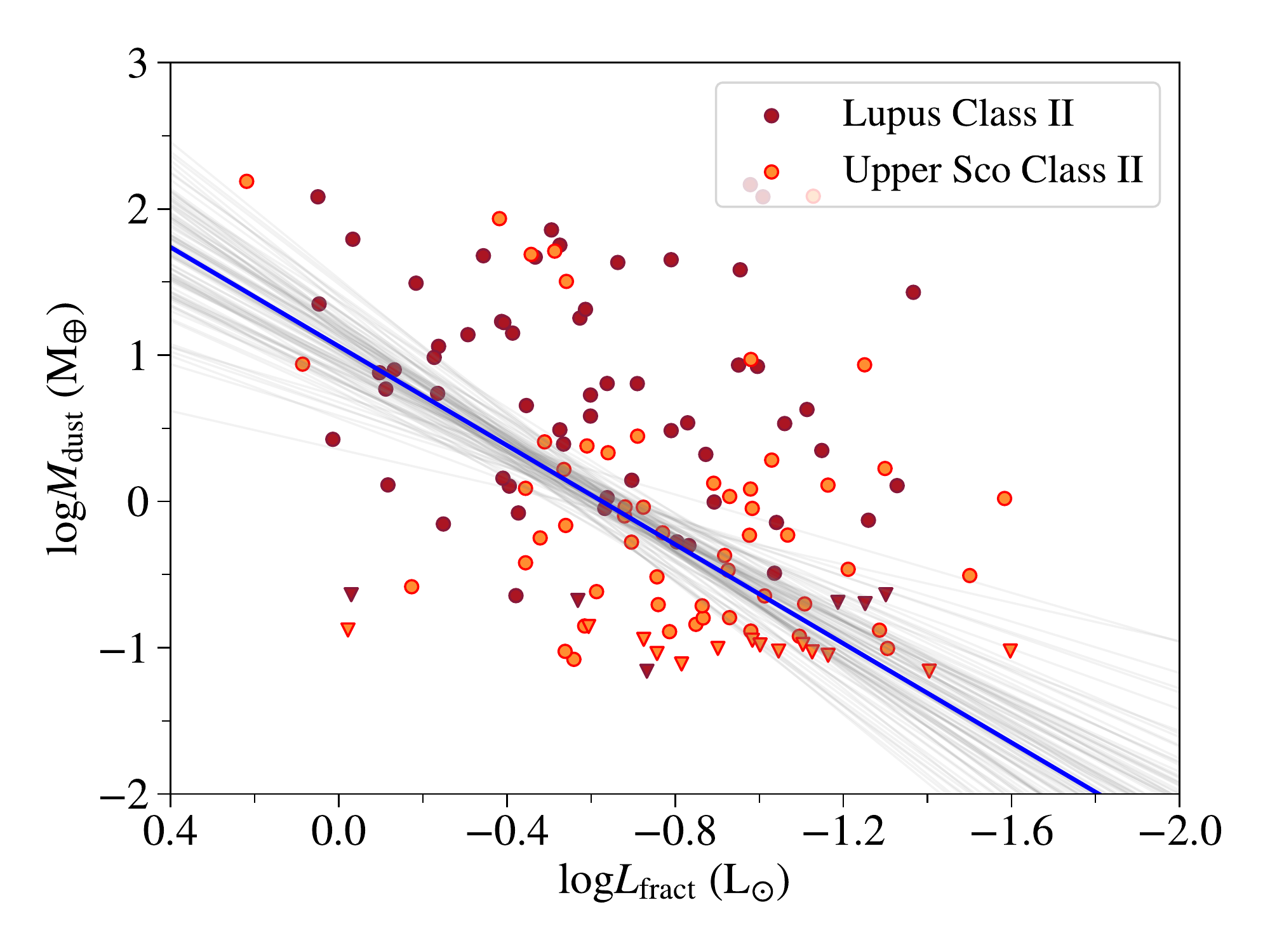}{0.5\textwidth}{(a)}
          \fig{macc_classII_2021_02_25.pdf}{0.5\textwidth}{(b)}
          }
 \caption{{\bf Left:} Dependence of $M_{\text{dust}}$ and $L_{\text{fract}}$ for the Class II disks from Lupus (in red) and Upper Sco (in yellow). The blue line shows the best-fit correlation between these two parameters using Bayesian analysis, but only a weak correlation ($r_{\rm corr}=0.44\pm0.11$) was found. {\bf Right:} Dependence of $\dot{M}_{\text{acc}}$ and $L_{\text{fract}}$ for the Class II disks for which accretion values were available (Class III disks are only plotted for illustrative purposes - they are not included in the fit). A stronger correlation ($r_{\rm corr}=0.68\pm0.31$) was found for these parameters. For both panels, round scatter points represent detections while downward facing triangles are non-detections. \label{fig:lfracfit_lfracmacc}}
\end{figure*}

 Bayesian analysis is used to check whether there is a statistical correlation between $L_{\text{fract}}$ and $M_{\text{dust}}$ for Class II disks in Lupus and Upper Sco. A linear fit in log-log space using the \textit{linmix} tool, which accounts for non-detections and uncertainties for both input variables \citep{Kelly_2007}, yields a correlation coefficient, $r_{corr}$, of $0.46 \pm 0.11$ (left panel (a) of Figure~\ref{fig:lfracfit_lfracmacc}), implying only a moderate correlation. For these Class II disks, as $L_{\text{fract}}$ decreases, there is only a mild corresponding effect on $M_{\text{dust}}$. Interestingly, the ring and transition disks or `structured disks' (marked separately in Figure~\ref{fig:lfractmdust}), have the highest dust masses and do not exhibit an observed decrease in dust mass between Lupus and Upper Sco. In our sample, 18/19 of the massive disks with $M_{\text{dust}}>25~\text{M}_{\oplus}$ are structured, and among these structured disks, we find a mean $M_{\text{dust}}$ of $62~\text{M}_{\oplus}$. Excluding the structured disks from the Lupus and Upper Sco samples results in the KME calculation results in a mean dust mass of $5.6 \pm 1.1~\text{M}_{\oplus}$ for Lupus and $1.0 \pm 0.2~\text{M}_{\oplus}$ for Upper Sco.

\paragraph{Class III regime}
 The disks in the Class III regime have millimeter dust masses that are up to ${\sim}2$ orders of magnitude lower than the Class II regime, with a mean $M_{\text{dust}}$ of $0.29 \pm 0.18~\text{M}_{\oplus}$ as calculated using KME. With our data, we cannot confirm whether the Class III disks display the same three orders of magnitude spread in dust mass as the Class II disks. The sensitivity of the millimeter data is insufficient to probe over such a range and thus contains many upper limits; in addition to this, the sample is non-uniformly constructed. The underlying Class III distribution may be more tightly constrained than the Class II distribution due to evolutionary processes, but we cannot ascertain this from these data.

 Furthermore, the decline in $M_{\text{dust}}$ between the Class II and III phases doesn't affect all disks equally: the structured disks appear to follow an alternate evolutionary track with a delayed mm-dust decrease. Notably, we identify, 2MASS J16083070-3828268 (a Lupus Class II disk) and Sz~91 (a Lupus Class III disk, classified according to the $\alpha_{\mathrm{Lada}}$ scheme) that have much lower fractional disk luminosities but mm-disk dust masses comparable to the majority of the structured disks. These two objects have $L_{\text{fract}}$ of $4.3\times10^{-2}$ and $7.2\times10^{-2}$ with $M_{\text{dust}}$ of 28.4 and $16.4~\text{M}_{\oplus}$, respectively. In comparison, the rest of the objects (Class II and III targets) within this same range of fractional disk luminosities have a mean $M_{\text{dust}}$ of $0.76 \pm 1.64~\text{M}_{\oplus}$. For most Class III objects, it is not possible to assess whether they are structured as the ALMA data do not have sufficient resolution (${\sim}$1"). However, it is very likely that these low values of $L_{\text{fract}}$ are simply the result of rare, very large dust cavities ($>$50 au) in late type stars: as both of these disks are still gas-rich and accreting \citep{vandermarel_2018} they are likely not evolved, but primordial with a large inner cavity.
 
 \paragraph{Debris disk regime}
 For disks with $L_{\rm fract} < 10^{-3}$, $M_{\text{dust}}$ decreases steadily with the fractional disk luminosity (see Figure~\ref{fig:lfractmdust}). The KME-calculated mean mass of the debris disk sample is $0.053 \pm 0.008~\text{M}_{\oplus}$. We emphasize that this sample is biased towards early type stars compared to the Class II and III samples (see Figure~\ref{fig:sptdist}).

\subsection{Accretion}
 Figure~\ref{fig:lfracfit_lfracmacc}b shows the accretion rates of Class II and III disks as a function of fractional disk luminosity. As for most Class III objects no accretion rates have been derived, and as the data points that are available are primarily upper limits (14/20), the plot is dominated by Class II disks and meant to illustrate the evolution in the Class II phase. The accretion rates of Class II disks in older regions (Upper Sco, TW Hya, and $\eta$ Cha) have a similar range despite their significantly lower millimeter dust masses, as noticed by \citet{Manara_2020}. Note that we use a ${\sim}30\%$ uncertainty in $\dot{M}_{\text{acc}}$ as the distance uncertainties are much lower in light of \textit{Gaia} parallax in comparison to previous studies, thus we adopt the typical error from \citet{Alcala_2017} rather than their total uncertainty that factored in an additional 23\% uncertainty from distance measurements. The \textit{linmix} fit retrieves a correlation coefficient of $0.68\pm0.31$, which is within error bars the same as the $L_{\text{fract}}$ and $M_{dust}$ correlation coefficient, although a potential stronger correlation cannot be excluded. The Class III objects visually appear to follow the decreasing trend of accretion rates with fractional luminosity, although the numbers are too small and non-uniform to confirm this. We do check for Class II sources that the correlation between $L_{\text{fract}}$ and $\dot{M}_{\text{acc}}$ is not simply based on a dependence on $L_{*}$ and find that the correlation coefficient between $L_{\text{fract}}$ and $L_{*}$ is ($r_{\rm corr}=0.53\pm0.12$; see Figure~\ref{fig:Lfract_Lstar_Macc} in the Appendix~\ref{appendix:Lfract_Lstar_Macc}).

\subsection{Disk lifetime}

\begin{figure*}
   \centering
   \includegraphics[width=\textwidth]{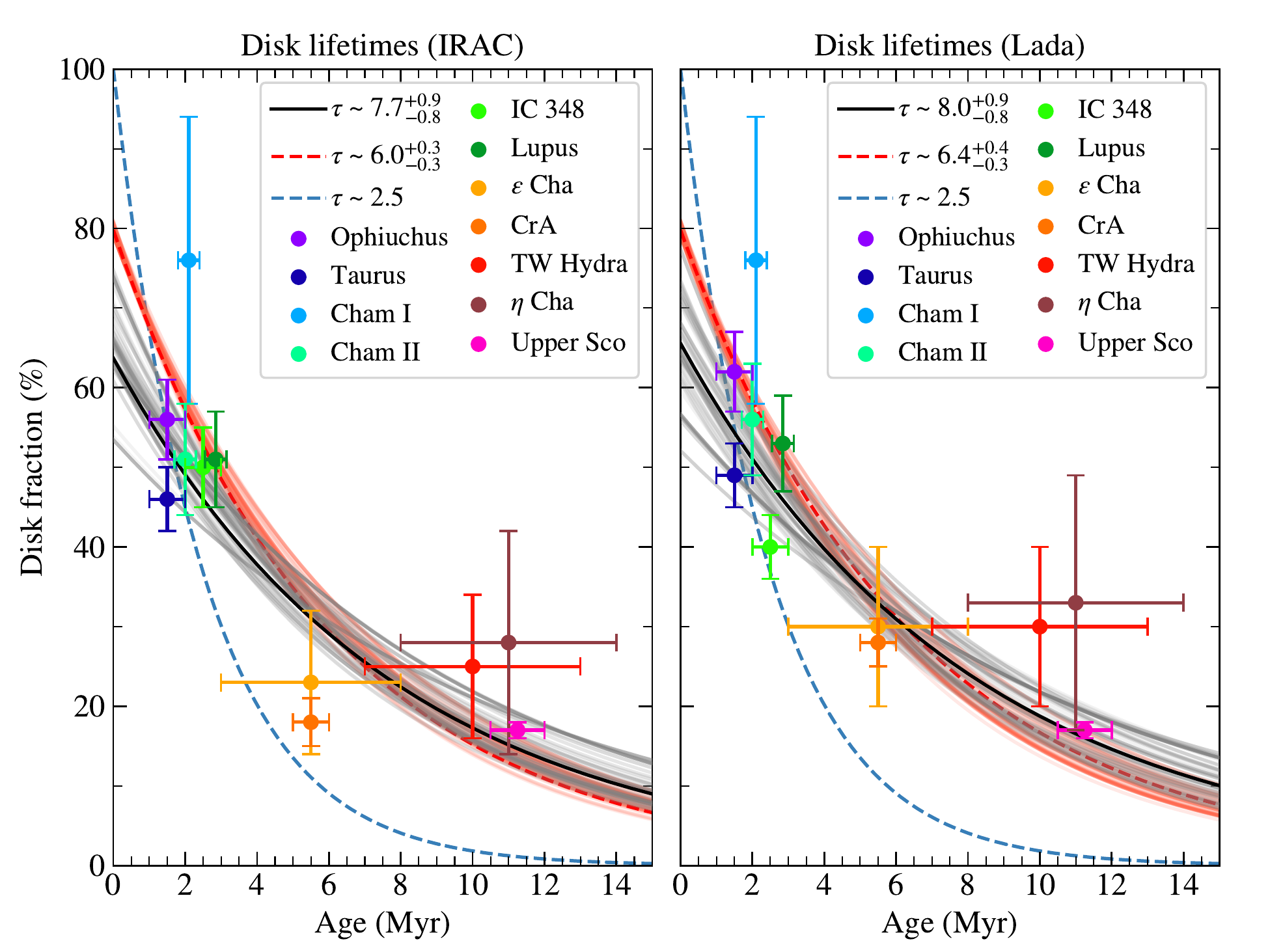}
        \caption{The disk fraction of each star-forming region as a function of age based on Gaia-constrained and updated star-forming region stellar membership  using the values from Table \ref{tbl:regiontable}. The left panel shows the disk fraction according to the IRAC classification, based on $\alpha_{\text{IRAC}}$; the right panel is according to the Lada classification, based on $\alpha_{\text{Lada}}$. The black solid line represents the free parameter best-fit exponential line for the timescales in each plot. The red dashed line represents the fixed parameter ($A=80\%$) best-fit exponential line. The transparent lines are random draws from the posterior distribution that resulted from the MCMC fitting. The blue dashed line shows the best fit according to \citet{Mamajek_2009} using a wider range of star-forming regions and older disk fraction values.}
        \label{fig:diskfrac}
\end{figure*}

 The star-forming regions' disk fractions in Table~\ref{tbl:regiontable} can be used in combination with their age to determine the average expected lifetime of Class II disks. Figure \ref{fig:diskfrac} shows the disk fraction as a function of age across star-forming regions, where the left panel is based on the IRAC classification, and the right panel is based on the Lada classification. For both of these disk fraction evolutionary trends, we calculate an expected exponential decay similarly to \citet{Mamajek_2009, Ribas_2014} of the form:
 \begin{equation}
    P_{diskfrac} = Ae^{-t/\tau} \label{eq:diskfrac},
 \end{equation}
 where $t$ is the age of the star-forming region in Myr as found in Table~\ref{tbl:regiontable} and both $A$ and $\tau$ are left as free parameters. The timescale $\tau$ is taken to be the characteristic timescale of infrared excess decay. The parameters are fit using the \texttt{emcee} package to compute the posterior distributions \citep{ForemanMacKey2013}. We perform the fit both with $A$ as free parameter and with $A$ fixed to 80\%, assuming an initial binarity fraction of 20\% as binarity reduces the disk fraction \citep{Kraus_2012}. A fit with $A$ fixed to 100\% does not converge. For the free $A$, we find characteristic time scales $\tau$ of 7.7$^{+0.9}_{-0.8} $~Myr for the IRAC disk fractions and 8.0$^{+0.9}_{-0.8}$ Myr for the Lada disk fractions, with an $A$ of $63.7^{+5.2}_{-4.6}$\% and 65.6$^{+5.0}_{-4.5}$\%, respectively. When $A$ is fixed such that the disk fraction at 0~Myr is 80\%, the $\tau$ values reduce to $6.0^{+0.33}_{-0.3}$ and 6.4$^{+0.4}_{-0.3}$ Myr respectively, but we stress that this is not an intrinsically better fit. We use the Bayesian Information Criterion (BIC) test to determine the statistically better fit. In both cases we find that the free $A$ fits results in lower BIC values, 30 (IRAC) and 33 (Lada) compared to the fixed $A=80\%$ fits for which we find BICs of 36 (IRAC) and 39 (Lada) suggesting that the free parameter fit is better suited to describing the characteristic disk lifetimes. In addition, the binarity fraction was derived as 20$\pm$5\% from a sample with a spectral type range between G0 and M4 \citep{Kraus_2012}, whereas our samples contain a wider range of spectral types and multiplicity is a function of stellar mass \citep{Moe_2019}. This might explain why a better fit was found with a lower $A$ value, essentially suggesting a higher binarity fraction than previously derived, or additional processes suppressing disk formation. This requires more detailed investigation of binarity in young clusters.
 
 The derived lifetimes are much larger than the $\tau$ values of 2-3 Myr as derived by \citet{Mamajek_2009,Williams_2011}. While a 2.5~Myr timescale from \citet{Mamajek_2009} suggests that only 5\% of protoplanetary disks still exist by an age of 7.5~Myr, our results indicate that in our chosen regions this would be on the order of ${\sim}20$~Myr. The quality of the fits suggests that this simple model does not fully represent the data. The two best-fit exponential decay models do still demonstrate that the expected disk fraction decreases with age but rather than turning into a completely diskless region in $\lesssim10$ Myr, we show that the objects retain optically thin disks, potentially evolved ones, for a much longer period of time, on the order of ${\sim}20$~Myr.

\section{Discussion}\label{sec:discussion}

\subsection{Disk evolution}\label{subsec:evolution}

 Our work demonstrates that $M_{\text{dust}}$ decreases as disks evolve but the disk dust clearing processes are more complicated than a simple linear decrease over time. The mm and $\mu$m-sized dust traced by $M_{\text{dust}}$ and $L_{\text{fract}}$ in disks as seen in Figure~\ref{fig:lfractmdust} provide a more detailed picture of how the disks evolve and how disk clearing processes remove dust of different sizes during the succeeding disk stages. It is important to note here that individual disks evolve on different timescales depending on their initial conditions, and that age is not necessarily a direct measure of disk evolution. We identify and label the three distinct stages in the disk evolution process at $L_{\text{fract}}$ values of ${\sim}10^{-1}$ and $10^{-3}$ that delineate the border between Class II-III disks and Class III-debris disks. The two dust tracers seem to indicate marked evolutionary differences between these disk evolution stages.
 
 The strong decline in $M_{\text{dust}}$ within Class II disks from Lupus (1-3 Myr) to Upper Sco (10.5-12 Myr) seen in Figure~\ref{fig:lfractmdust} with the exception of structured disks can be explained by radial drift and dust growth \citep{Weidenschilling_1977} rather than dissipation. Structured disks (transition and ring disks) do not appear to be affected by radial drift and retain their mm-dust, as expected from their proposed origin as pressure bumps \citep{Pinilla_2018,Pinilla_2020}. This scenario is further supported by the high accretion rates in Upper Sco suggesting limited gas dissipation \citep{Manara_2020} and the decrease in dust disk radius of the older Upper Sco disks \citep{Hendler_2020} which is a direct sign of inward radial drift. A recent work on disk evolution including both radial drift and low viscosity of $\alpha_{v}=10^{-3}$ reproducing these observational results \citep{Sellek_2020} shows the importance of including dust evolution in evolutionary disk models, this is further supported by modeling results by \citet{Appelgren_2020} on planetesimal formation in the inner disk. Also, a study of disk gas radii by \citet{Trapman_2020_spreading} suggests low $\alpha_{v}=10^{-4} - 10^{-3}$ values. The spread in the initial disk and stellar conditions and the dependence of drift on stellar mass \citep{Pinilla_2013, Zhu_2018} results in a wide range of $L_{\text{fract}}$ and $M_{\text{dust}}$ values. This results in a large $M_{\text{dust}}$ distribution within the Class II phase which doesn't homogeneously evolve into Class III disks. We will further discuss the structured disks in Section~\ref{sec:structured}, but for the moment we focus on the majority of the Class II disks in our sample, whose $M_{\text{dust}}$ decreases before gradually dissipating and evolving to Class III objects.
 
 In this disk evolution scenario, where the mm-dust grains (partially decoupled from the gas) experience radial drift and rapid dust growth, $\mu$m-sized dust grains are expected to remain well coupled with the gas, and their evolution could be traced by the $\dot{M}_{\text{acc}}$. The gas will then dissipate later in the Class III stage  \citep{Alexander_2014,Ercolano_2017}. As such, we would expect a stronger correlation between $L_{\text{fract}}$ and $\dot{M}_{\text{acc}}$ compared to $L_{\text{fract}}$ and $M_{\text{dust}}$. As the difference between the correlation coefficients from Figure~\ref{fig:lfracfit_lfracmacc} is inconclusive, further observations to measure $\dot{M}_{\text{acc}}$ and $M_{\text{dust}}$ of complete Class II and III disk samples are necessary to reveal such a difference.
 
 Disk dissipation carries disks into the Class III phase. In this phase, the disks are now optically thin in the infrared, but critically, $M_{\text{dust}}$ has already decreased, as 97.5\% of the disks have $M_{\text{dust}} < 1.2~\text{M}_{\oplus}$. It is important to note here that previous works comparing dust masses of protoplanetary disks with debris disks reported a significant decrease in dust mass of 2-3 orders of magnitude between these two populations \citep{Wyatt_2008,Panic_2013,Hardy_2015}, which they interpret as rapid photoevaporative clearing \citep{Clarke_2001}. However, these studies mostly included pre-ALMA measurements of protoplanetary disks, which did not have the sensitivity of the more recent ALMA surveys \citep{Ansdell_2016,Barenfeld_2016,Williams_2019,Lovell_2021}. The current data reveal that even in the protoplanetary disk phase the mass in millimeter grains has already declined to levels comparable to those measured in debris disks, although the debris disk sample, due to the aforementioned biases (see Section~\ref{sec:results}), is unlikely to represent the same population as these protoplanetary disks. This decrease in $M_{\text{dust}}$ is attributed to radial drift and rapid dust growth rather than dissipation. The predicted rapid photoevaporative clearing process \citep{Clarke_2001} may well be beyond the sensitivity of current observations. 
 
 The continuous dissipation within Class II disks from the overall younger Lupus to the older Upper Sco region demonstrates that radial drift has been efficiently removing mm-dust, decreasing $M_{\text{dust}}$, but it is only from the transition to the Class II to III regimes that the disk dissipation becomes clearly visible as the infrared emission becomes optically thin and $L_{\text{fract}}$ decreases further. Radial drift, on the other hand, has already greatly impacted the disks; it explains why within the Class II regime the mean disk mass of the non-structured disks is $5.6 \pm 1.1~M_{\oplus}$ for Lupus, while for Upper Sco it is $1.0 \pm 0.2~{M}_{\oplus}$, similar to dust evolution predictions over 1-10 Myr for a drift-dominated disk \citep{Pinilla_2020}. However, a trend with dust evolution in $M_{\text{dust}}$ and $L_{\text{fract}}$ cannot be identified in the Class III regime due to a large number of upper limits in that sample. It is thus not possible to distinguish the Class III remnants from older, more drift-affected disks in Upper Sco from those in younger regions. On the other hand, it is certainly plausible that most disks evolve along such an evolutionary drift-dominated path, first mainly decreasing in $M_{\text{dust}}$ through radial drift and also through dust growth, followed by declining $L_{\text{fract}}$ through dissipation into the optically thin infrared regime.
 
 Class III disks continue to evolve with disk dissipation mechanisms including photoevaporation, followed by radiation pressure and Poynting-Robertson drag \citep{Williams_2011,Wyatt_2015}, shifting $L_{\text{fract}}$ downwards. In essence, by the time the disks evolve from Class II disks into and through the Class III evolutionary stage, the $\dot{M}_{\text{acc}}$ will eventually drop below the photoevaporation threshold and the latter starts to dominate \citep[the so-called `UV-switch',][]{Clarke_2001}, the disk is rapidly drained inside-out \citep{Alexander_2014,Sellek_2020}. Only the largest solid components (boulders and planetesimals) which are completely decoupled from the gas remain \citep{Wyatt_2015}. If there are sufficient planetesimals and efficient stirring, these large bodies can produce smaller, second-generation dust particles through collisions and gradually reach a dust grain size that can be efficiently removed from the system by radiative forces \citep{Wyatt_2008,Holland_2017,Hughes_2018}. As debris disks generally exhibit cold dust belts at large orbital radii of tens of au \citep{Holland_2017,Matra_2018}, this requires that these belts of planetesimals are located at such large distances as well.
 
\begin{figure}[!ht]
   \centering
   \includegraphics[width=0.5\textwidth]{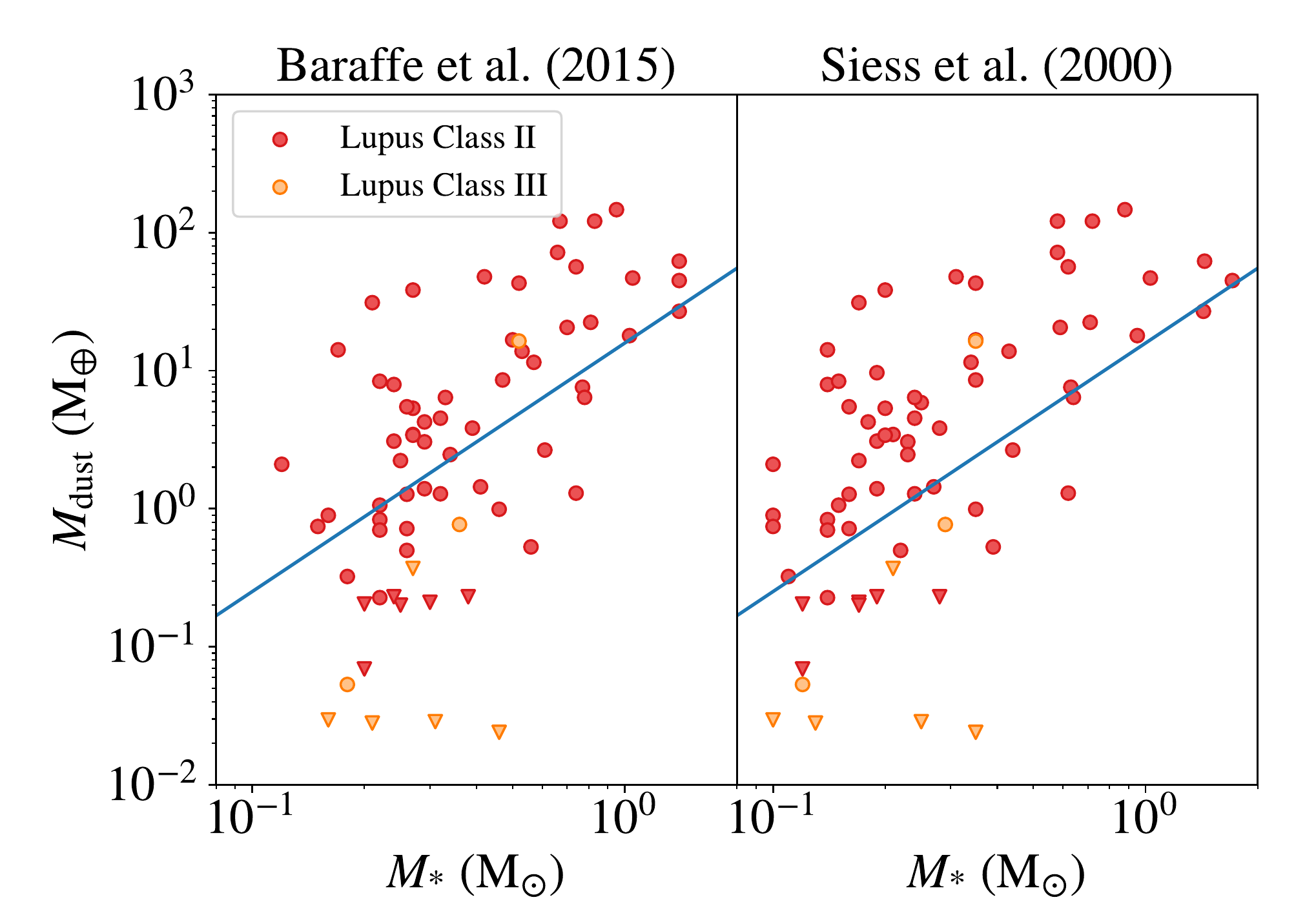}
      \caption{The $M_{\text{dust}}$-$M_*$ plot of Lupus Class II and III targets, following Figure 11 in \citet{Lovell_2021}, but only including Lupus members according to the Gaia assessment study by \citet{Luhman_2020_lupus}. The rapid dust dissipation claimed by \citet{Lovell_2021} is no longer visible in this plot.}
         \label{fig:mdustmstar_lupus}
\end{figure}

 \citet{Lovell_2021} present a different scenario to connect their Class III disk observations to debris disks. They suggest that planetesimal belts found in debris disks are likely already formed after 2 Myr, due to rapid dust mass evolution between Class II and III stage. However, according to recent \textit{Gaia}-based membership studies of Lupus \citep{Luhman_2020_lupus,Galli_2020_lupus}, most (24/30) of \citet{Lovell_2021} targets do not appear to belong to the Lupus star-forming region itself, but are likely part of older surrounding Sco-Cen regions (e.g. UCL aged ${\sim}$16~Myr \citep{Pecaut_2012}), putting these timescales into question. In particular, the strong drop in dust mass between Class II and Class III Lupus targets in the $M_{\text{dust}}$-$M_*$ plot leading to their conclusion of early rapid dust dissipation, seen in their Figure~11 \citep{Lovell_2021}, is no longer visible when non-members are excluded, see our Figure~\ref{fig:mdustmstar_lupus}. In particular, one could conclude that the observable millimeter dust masses drop from $\gtrsim25M_{\oplus}$ for the Class II disks in Lupus around Solar mass stars ($\lesssim0.04M_{\oplus}$) to disks in Upper Cen over $\sim$16 Myr due to planetesimal formation. In comparison, the handful of Class III disk detections in young 1-2 Myr clusters from Table \ref{tbl:mastertable} suggest observable dust masses of $\lesssim0.4M_{\oplus}$ in this regime instead, well above the typical debris disk dust mass. However, due to the incompleteness of the Class III samples it is not possible to draw strong conclusions about the planetesimal growth timescales here.

 \subsection{Ring and transition disks}\label{sec:structured}
 Figure \ref{fig:lfractmdust} shows that the majority of our sample of massive disks ($\gtrsim 25~M_{\oplus}$) are structured; they have been identified in millimeter images as large-scale transition disks ($>20$~au inner dust cavity) or ring disks (one or multiple dust gaps) by \citet{Andrews_2018,Cieza_2021}. The gaps and cavities are thought to be carved by sub-Jovian and Jovian planets, where pressure bumps at the edge trap the millimeter dust in a ring-like structure: a dust trap \citep{Pinilla_2012,vandermarel_2016}. Although the identification of transition and ring disks is limited by the spatial resolution of the observations, it is remarkable that most structured disks are found in the high disk dust mass regime. Furthermore, the occurrence of structured disks has been shown to increase with stellar mass \citep{vanderMarel_2021}, but this is not directly evident in our Class II sample due to the large absolute number of K and M stars. The consequences with respect to spectral type are discussed in the next section.

 The decrease in $M_{\text{dust}}$ with $L_{\text{fract}}$ in the Class II phase, as discussed in Section \ref{subsec:evolution}, caused by radial drift does not appear to affect the structured disks: whereas the majority of the dust masses in Upper Sco disks lie well below those of Lupus, the structured disks' $M_{\text{dust}}$ values are similar between these two regions. The structured disks appear to follow an alternate evolutionary track with a delayed decrease in $M_{\text{dust}}$ similar to the outlier disks from \citet{Ansdell_2020}. Dust traps prevent these disks from undergoing radial drift-dominated mm-dust evolution while the $\mu$m-sized dust is still being gradually dissipated, thus they retain their high dust mass \citep{Pinilla_2020, Sellek_2020,Cieza_2021,vanderMarel_2021}. Within the dust traps, at the edge of gaps and cavities, dust grains continue to grow, although likely at a slower rate than in the inner part of a radial drift dominated disk, considering the lack of decrease in $M_{\text{dust}}$ in structured disks. Dust evolution simulations by \citet{Pinilla_2020} indeed show only a small decrease in $M_{\text{dust}}$ in the models with pressure bumps, compared to smooth models, consistent with this scenario. This process results in locking up the mm-dust grains in larger bodies at larger radii and can be successful in forming well-populated planetesimal belts. The dust can grow to form planetesimals by the streaming instability in the dust traps \citep{Johansen_2007}; this process has been suggested to explain the observed characteristics of rings in DSHARP disks \citep{Stammler_2019}. Essentially, this results in two separate evolutionary pathways for structured and non-structured disks, such as previously suggested for Herbig disks \citep{Garufi_2017} and T Tauri stars in Lupus \citep{vandermarel_2018}. A connection between structured disks and debris disks was also suggested in a recent work by \citet{Cieza_2021}.
 
 The disk dissipation through viscous accretion continues in the structured disks until photoevaporation takes over, quickly removing the remaining gas and small dust grains, except for dust that has grown into boulders and planetesimals as they have become decoupled from the gas \citep{Pinilla_2020}. Drift-dominated disks may also have formed boulders and planetesimals in the inner part of the disk \citep{Pascucci_2016}. It is unclear whether the separation between structured and unstructured disks is reflected in the Class III disk population since the ALMA observations of the Class III objects generally have a spatial resolution that is too poor (${\sim} 1"$) to resolve any structure. However, two disks are detected with ALMA and located at the higher end of the Class III $M_{\text{dust}}$ regime with relatively low $L_{\text{fract}}$ values, 2MASS J16192923-2124132 and 2MASS J04192625+2826142. Based on the literature, these appear to possibly be structured disks with large cavities: 2MASS J16192923-2124132 is found to have a best fit inner disk radius of 71~au based on ALMA visibility modeling \citep{Lieman-Sifry_2016} and 2MASS J04192625+2826142 has a cavity radius of 150~au based on SED modeling \citep{vandermarel_2016}. Similar to the Class II structured disks, these are located at the upper range of the Class III dust masses.

 In the debris disk phase, dust emission is detectable in deep imaging because of the proximity and isolated positions of the disk hosts in the sky. The debris disks targeted by \citet{Holland_2017} consist of cold dust belts at ${\sim}10-200$~au (see $R_{\text{BB}}$ column in their Table 3) which are the result of the collisional evolution of planetesimals under the influence of stirring. The planetesimals are thus constrained to large radii, which can only be understood if they are formed in  pressure bumps at large orbital radii in the protoplanetary disk phase (as illustrated by the topmost arrow in Figure.~\ref{fig:lfractmdust}). Therefore, we hypothesize that only one of the proposed evolutionary paths above result in debris disks: the observed debris disks are the outcomes of structured disks, whereas drift-dominated disks evolve into near diskless stars (diskless stars at the current observational capabilities). In the absence of significant traps such as those presented by \citet{Pinilla_2020} there simply is no mechanism by which radial drift and dust growth will allow for a significant planetesimal belt formation, thus there cannot be significant second-generation dust production as observed in debris disks dust belts. 
 
 More quantitatively, the connection between Class II structured disks and debris disks can be seen through the similar radius location for their outer rings/cavity inner edges and the debris disk dust belts. Comparing the cavity radii $R_{\text{cav}}$ sample of 38 transition disks observed with ALMA from \citet{Francis_2020} to the radii of the dust belt $R_{\text{BB}}$ of the 49 detected debris disks from \citet{Holland_2017}, we find similar median values, that is $47\pm44$~au and $49\pm156$~au, respectively. The range of radii for 95\% of the transition disk cavities is $23 < R_{\text{cav}} < 188$~au while for 95\% of debris disk that is $12 < R_{\text{BB}} < 248$~au thus demonstrating the similarities present in biased samples of structured disks and currently observed debris disks. We note, however, that debris disk radii can be 1 to 2.5 times the $R_{\text{BB}}$ value, due to the fact that the dust grains are not perfect blackbodies \citep{Booth_2013}. 

 The hypothesized connection between structured disks and debris disks requires further investigationa and modeling which is beyond the scope of this work. However, \citet{Jiang_2021} show in their recent modeling work how structured disk rings (e.g., as seen in DSHARP) can be long lived and the precise location where massive planetesimal belts form and become debris disks.  
 
 \subsection{Comparison with spectral types}
 \label{subsec:spt}

\begin{figure}[!ht]
   \centering
   \includegraphics[width=0.5\textwidth]{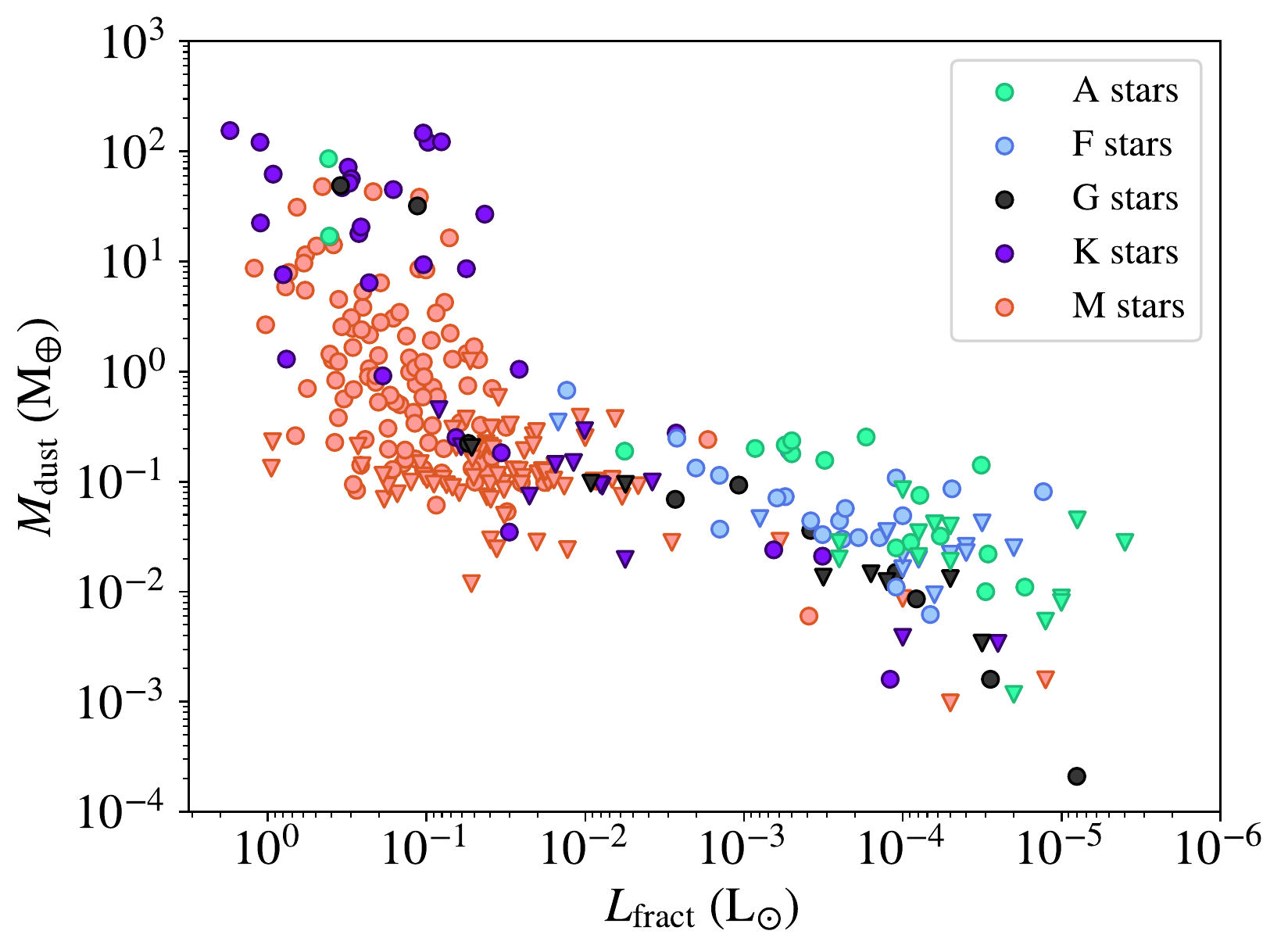}
      \caption{Fractional disk luminosity vs mm-dust disk mass, such as Figure \ref{fig:lfractmdust}, but color-coded by spectral type. The debris disk sample is clearly biased towards early type stars compared to the general Class II and III samples.}
         \label{fig:lfractmdust_spt}
\end{figure}

 A major uncertainty in our comparison is the difference in spectral types between our Class II and Class III disks, and the debris disk sample. Debris disks are more commonly detected around early-type stars \citep{Matthews_2014,Sibthorpe_2018,Hughes_2018} and the SONS sample exacerbates this bias by exclusively targeting known disks detected in the infrared (and extrapolated to be detectable in the sub-mm by the JCMT). Hence its stellar mass distribution is very biased towards early-type stars (Figures~\ref{fig:sptdist} and~\ref{fig:lfractmdust_spt}): the SONS sample is dominated by A and F-type stars, with a fair number of G and K-type stars, and only four M-type stars. 
 
 The key point in the comparison between structured disks with rings at large radii and debris disks is that the protoplanetary disks represent a \emph{complete} population (or IMF) dominated by K and M stars, whereas the debris disk sample does not. As the occurrence of structured disks is higher for early type stars \citep{vanderMarel_2021} but early type stars are intrinsically also rarer in the IMF, our sample of structured disks may appear to be dominated by K and M stars, but relative to the total disk population, it is not. In contrast, the debris disk sample encompasses a range of 10 Myr -- 6 Gyr and thus contains the products of hundreds of clusters, and as the structured disks are found more commonly around early type stars, the debris disk sample from \citet[][]{Holland_2017} would naturally be dominated by early type stars as well if their progenitors were structured disks. The observed occurrence rate as function of stellar mass for debris disks and structured disks is indeed similar \citep{vanderMarel_2021}.

 One caveat in the comparison of occurrence rates is that debris disks around M stars are particularly scarce down to existing detection limits \citep{Luppe_2020}. We suggest that there may be inherently fewer debris disks around M stars due to their possible lack of formation; in addition, the effects of rapid collisional evolution and stellar winds to deplete and cut off the dust size distribution render those that do exist particularly difficult to detect \citep{Plavchan_2005}. The remaining parameter space where disks around M stars could be prevalent, both being very cold and very small, is hard to explore and as of yet would not match the current definitions of the typical cold debris disk. 

 This connection implies that most of the protoplanetary disks lack the distinct and large structures that can produce planetesimal belts at large radii which we expect to evolve into cold debris disks. Furthermore, considering the spectral type differences, the dust mass distribution of the full Class II and III population should not be directly compared with the debris disk dust mass population as a whole.
 
 Considering Figure \ref{fig:Lfract} on a quantitative level, only the dust mass of K type stars can be compared directly: the dust mass of structured disks in the Class II phase of $\gtrsim 25~M_{\oplus}$ drops to $\sim0.01M_{\oplus}$ for debris disks around K type stars, but these are small sample statistics and more data are required to draw meaningful conclusions. 
 
 \subsection{Comparison with ages}
 \label{subsec:ages}
 
  \begin{figure*}
   \centering
   \includegraphics[width=\textwidth]{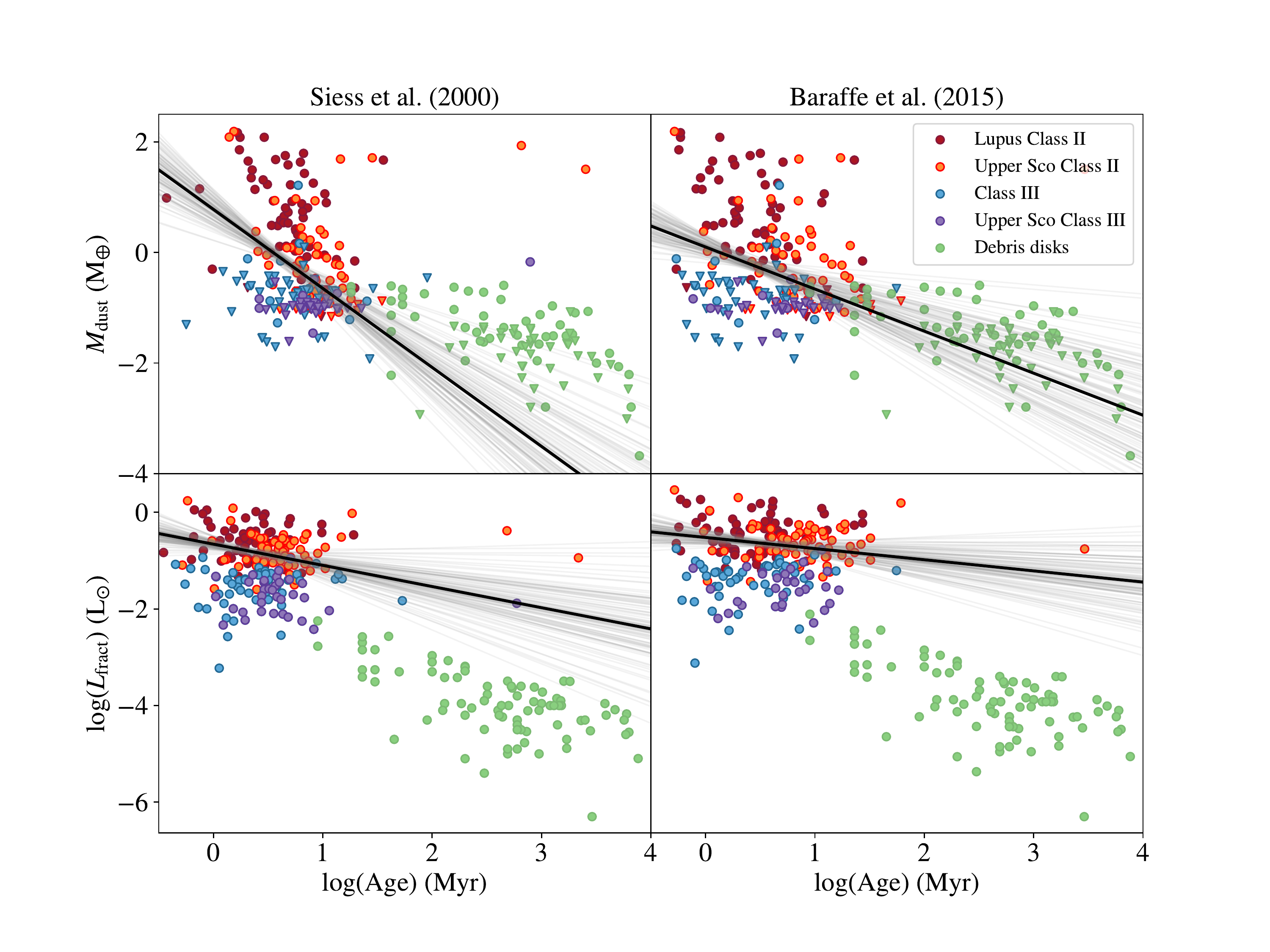}
      \caption{Age using different evolutionary models. Plot colours and symbols are identical of Figure~\ref{fig:lfractmdust}. For Class II and Class III disks, the dependence of the $L_{\text{fract}}$ and $M_{\text{dust}}$ are explored in relation to age of individual targets derived from two models \citep{Siess_2000,Baraffe_2015}. The black line shows the best-fit correlation between each pair of parameters using Bayesian analysis. No correlation was found, see the text for the correlation coefficients.}
         \label{fig:age}
\end{figure*}

 Many disk evolution studies connecting protoplanetary and debris disks aim to relate the individual ages of disks obtained from stellar evolutionary isochrones with either $\dot{M}_{\text{acc}}$ or $M_{\text{dust}}$ to study evolutionary trends \citep[e.g.][]{Cieza_2007,Wahhaj_2010,Hardy_2015}. The age of protoplanetary disks is an unreliable comparison variable due to the current uncertainty in the individual estimates \citep{Soderblom_2014}. \citet{DaRio_2014} demonstrated that tracking the relationship of $\dot{M}_{\text{acc}}$ against individual ages could be spurious due to stellar parameter uncertainties. 

 In Figure~\ref{fig:age}, we explore the possible $M_{\text{dust}}$ and $L_{\text{fract}}$ trends compared to age using two different evolutionary isochrone models from \citet{Baraffe_2015} and \citet{Siess_2000}, using the spectral types from the literature, re-scaled luminosities, and the debris disk ages as derived in \citet{Holland_2017}. We use the \textit{linmix} tool to obtain a linear fit in log-log space between these parameters, to search for a correlation. For the protoplanetary disks (Class II and III), we find no correlation between the age and $M_{\text{dust}}$ ($r_{corr}$= $-0.26 \pm 0.08$ \citep{Baraffe_2015} and $-0.32 \pm 0.08$ \citep{Siess_2000}) and between age and $L_{\text{fract}}$ there is no correlation either ($r_{corr}$= $-0.17 \pm 0.11$ \citep{Baraffe_2015} and $-0.26 \pm 0.10$ \citep{Siess_2000}) although it is expected that generally, both $M_{\text{dust}}$ and $L_{\text{fract}}$ should decrease as disks become older. Visually, in Figure~\ref{fig:age}, we can see large overlaps and no explicit trends; whereas, at least, the Upper Sco disks, with older ages expected to be on the order of $10.5-12~\text{Myr}$ \citep{Luhman_2020_usco}, should have been distinguishable on this plot, but they are not.

 It is only within the debris disk sample that the age and $L_{\text{fract}}$ correlate \citep{Holland_2017}, which can be understood since the collisional evolution regenerating the dust in debris disks depletes the parent body population over time \citep{Wyatt_2008}. Therefore, evolutionary stages and their trends should be examined without trying to force imperfect age fits onto individual protoplanetary disks. Besides, the age of individual YSOs and the disk properties are expected to also vary based on the initial stellar host properties \citep{Mulders_2017}, the dust clearing processes' efficiencies \citep{Pinilla_2020}, and the star-forming region environment's UV impact \citep{Facchini_2016,Haworth_2017}. 

 Previous studies connecting debris disks with protoplanetary disks \citep{Cieza_2007, Wahhaj_2010,Hardy_2015} have typically targeted WTTS as a stage in between CTTS and debris disks. Although both WTTS and Class III objects are considered evolved protoplanetary disks, they are technically not the same, as explained in Section \ref{sec:accretionrates}. 
  
 These studies generally find lower disk fractions, $L_{\rm fract}$, $\alpha_{\text{IRAC}}$, and $M_{\rm dust}$ values for WTTS compared to CTTS, consistent with our findings for Class III disks. Using stellar isochrones they also find the average individual ages of WTTS to be located somewhere in between those of CTTS and debris disk hosts. Nevertheless, these have a large spread, as expected from the large uncertainties in this method. Lastly, it is very hard to know the collisional age of a debris disk as dust formed by planetesimal collisions and collisional cascades can sustain disks for several 10s of Myr \citep{Jackson_2012,Kenyon_2005}. Another complicating factor is that it seems plausible that "dormant" disks of planetesimals just await one large collision to sustain a fresh cascade, and while that is more probable in a young disk, it is not impossible in an old one \citep{Wyatt_2008}. This makes it difficult to establish timelines of when collisions have taken place in comparison to the original protoplanetary disk evolution and dissipation. Therefore, it remains a question of whether WTTS are truly older than CTTS, or simply more evolved. The $L_{\rm fract}$ parameter provides a more robust parameter for studying the disk evolution process. 

\subsection{Disk lifetime}
 The high disk fractions in star-forming regions from 1-12~Myr demonstrate that the disks are expected to be observable for several millions of years in the infrared, unlike the millimeter dust. The disk fractions are similar for both the IRAC and Lada classifications, regardless of the infrared wavelength range limits. The initial disk fractions in the young regions are not 100\%, but this is likely caused by binarity: close binaries ($<40$~AU) limit the formation of a disk or increase its dissipation process, which has been shown to lower the disk fraction in Taurus by at least 15-20\% \citep{Kraus_2012}. Similarly, close binaries ($<30$~AU) have been observed to be less likely to host a millimeter dust disk \citep{Harris_2012} and when they do the disks are smaller \citep{Akeson_2019}. Binarity studies for other regions are incomplete and we cannot confirm whether the same fraction applies to all regions, but it is plausible that all disk fractions will increase by at least 20\% if only single stars are included such as shown by \citet{Kraus_2012}. The best fit for the timescale for the disk fractions as a function of age in Figure \ref{fig:diskfrac} demonstrate that the infrared excesses from the $\mu$m-sized dust persist for longer time periods than previously thought \citep[e.g][]{Hernandez_2007,Mamajek_2009,Fang_2013,Ribas_2014} which we explain as being due to the consideration of UV environment. In these previous works, the characteristic timescales for infrared excess decay are on the order of 2-3 Myr \citet{Mamajek_2009}, or ${\sim}4$~Myr evaluated by \citet{Fang_2013} for sparse associations and depending on the wavelength used these can be between 2 to 6 Myr \citep{Ribas_2014} but without examining the UV field's impact. Instead, we find characteristic timescales in low UV nearby regions of ${\sim}8$~Myr depending on the disk fraction classification used with a free $A$ parameter. This corresponds to longer-lived infrared excesses and extended timescales for $\mu$m-sized dust grains in disks. 
 
 In both plots in Figure~\ref{fig:diskfrac}, the high disk fractions that still exist for TW Hydra, $\eta$ Cha, and Upper Sco for regions with ages on the order of ${\sim}10$ Myr demonstrate that the disk phase exists for longer than previously estimated. The long-lived $\mu$m-sized dust disks are consistent with the results from \citet{Manara_2020}, who find that $\dot{M}_{\text{acc}}$ values may still be generally higher than previously expected within older star-forming regions such as Upper Sco; there is still sufficient small dust and gas for significant accretion. However, note that the 36 Upper Sco targets with $\dot{M}_{\text{acc}}$ from \citet{Manara_2020} are an incomplete sample of the 1712 Upper Sco YSOs, thus these trends are suggestive rather than definitive statements. The longer disk lifetimes may even be able to explain the existence of $>$20 Myr old accretors around M dwarfs \citep{Murphy_2018, Silverberg2020}, although these have not been studied systematically yet and could be outliers. As the mm-dust decreases quickly due to radial drift, the $\mu$m-sized dust could be only gradually dissipating until the internal photoevaporation mechanism dominates the accretion rate and quickly drains the entire disk \citep{Alexander_2014,Sellek_2020}. More importantly, the longer infrared disk lifetime is consistent with a slower gradual disk dissipation process as illustrated in Figure~\ref{fig:lfractmdust} and thus a lower viscosity $\alpha_{v}$ on the order of $10^{-3}$ as used by \citet{Sellek_2020}. 
 
 Interestingly, some of our disk fraction estimates are similar to the disk fraction values used in previous works for the overlapping star-forming regions: it is thus not particularly the removal of background objects by the \textit{Gaia} membership assessment that results in a longer disk lifetime. One key difference is our selection of star-forming regions in relatively low UV environments, unlike previous disk fraction studies that included regions such as $\sigma$ Ori, $\lambda$ Ori, 25 Ori, Ori 1A \& B, NGC 2362, and NGC 7160 \citep{Hernandez_2007,Mamajek_2009,Ribas_2014}. These regions include O/B stars and are thus subject to strong UV field radiation \citep{Briceno_2005,Briceno_2007,Dahm_2007,Walter_2008,Bouy_2009}. The disk fraction has been observed to decrease close to the central OB stars in Orion, Cygnus OB2, and NGC 2024 \citep{Mann_2014,Guarcello_2016,vanTerwisga_2020}. In an analysis of Pismis~24, \citet{Fang_2012} demonstrate how disk fractions decrease in the presence of massive O stars within the region. Additionally, ALMA observations of protoplanetary disks in $\sigma$ Ori by \citet{Ansdell_2017} demonstrate that disk dust masses also decrease the closer the YSOs are to the central O9 star; this has also been similarly demonstrated by \citet{Eisner_2018} with the disk flux emission of Orion objects as a function of distance to $\theta^1$ Ori C, an O6 and B0 binary. The observations are supported by theoretical and modeling studies that show external photoevaporation impacts on the evolution of dust, amount of dust, and the protoplanetary disk lifetimes \citep[e.g][]{Scally_2001,Adams_2004,Facchini_2016,Haworth_2018,Winter_2018,Winter_2019,Sellek_2020_photoevaporation,Winter_2020}. Therefore, mixing both high UV and low UV star-forming regions in a single sample affects the resulting disk lifetime that can be obtained from these plots. The disk dissipation time scales could thus be longer in nearby star-forming regions where there are no strong external UV photoevaporation effects on the disks. The occasional B star (primarily late B-type) present in Taurus and Upper Sco, for example, are expected to have negligible impacts in comparison to the numerous O and early B-type stars in the aforementioned strong UV irradiated regions. In context, \citet{Trapman_2020_spreading} computed a median $F_{\mathrm{UV}}$ radiation of ${\sim}43$~G$_0$ for Upper Sco, which is one to two orders of magnitude lower compared to regions with strong UV field radiation, e.g., $\sigma$ Ori has an $F_{\mathrm{UV}}$ of $300\leq G_0 \leq1000$ \citep{Mauco_2016} and Cygnus OB has an $F_{\mathrm{UV}}$ on the order of a few thousand $G_0$ \citep{Guarcello_2016}.
 
 It is worth noting that we have not conducted an in-depth wavelength-dependent disk fraction assessment as presented by \citet{Ribas_2014}. Instead, we use two different wavelength ranges from 2.2 to 24 $\mu m$ ($\alpha_{\rm Lada}$) and 3.6 to 8.0 $\mu m$ ($\alpha_{\rm IRAC}$). These two wavelength ranges probe different parts of the disk; while the Lada classification covers a larger radial extent of the disk, the IRAC classification pertains to the inner couple au of the disk.
 
 A possible limitation to our disk lifetimes is that the disk fraction estimates for Ophiuchus are low, $P_{\text{diskfrac - IRAC}}$: $56 \pm 5$\% and $P_{\text{diskfrac - Lada}}$: $62 \pm 5$\%, compared to previous calculations; \citet{Canovas_2019} report a disk fraction based on 48 targets of 85\%. The stringent selection effects performed by \citet{Canovas_2019} on their Ophiuchus sample using solely \textit{WISE} data removed a majority of their disks. For Ophiuchus, we rely principally on the higher-quality \textit{Spitzer} IR data \citep{Avenhaus_2012} (see Appendix~\ref{appendix:regions}), thereby we deem it not mandatory to perform an intensive data selection for Ophiuchus.
 
  Another limitation within our determination of the disk lifetimes is the uncertainty of the region ages \citep[e.g.,][]{Bell_2015}. While the ages of the older regions, e.g. TW Hya and $\eta$ Cha, are fairly certain \citep{Bell_2015}, Upper Sco is an older region of contention which has earned a 5-12~Myr age range in the literature \citep{Preibisch_2002, Sartori_2003, Preibisch_2008, Slesnick_2008, Pecaut_2012} with more recent age descriptions favouring ${\sim}10\pm2$~Myr \citep{David_2016, Feiden_2016, Pecaut_2016, David_2019, Luhman_2020_usco}. The initial range of ages for Upper Sco includes, sample, distance, and model-based caveats and uncertainties. As such, different and previously incomplete stellar samples on which ages for Upper Sco have been determined thus require careful examination: see \citet[e.g.][]{Feiden_2016,Luhman_2020_usco} for further discussion on Upper Sco's age range and some of the past literature's shortcomings. If we were to consider a larger range of reported ages for Upper Sco across the literature from the past 20 years, resulting in an age range of from 5-12 Myr \citep{Preibisch_2002, Sartori_2003, Preibisch_2008, Slesnick_2008, Pecaut_2012, David_2016, Feiden_2016, Pecaut_2016, David_2019, Luhman_2020_usco}, we find that disk lifetimes with a free $A$ parameter could decrease to a range of 5-6~Myr, although we stress that many of the earlier studies underestimated the number of Upper Sco members and lacked \emph{Gaia} data to measure distance and membership. 
  
 In addition, the limitations of age determinations from pre-main sequence evolutionary isochrones is presented in a series of papers by \citet{Bell_2012,Bell_2013,Bell_2014} and highlights the issue in the frequent underestimation of the pre-main sequence stellar ages by factor of 2-5 compared to the more evolved main-sequence stars of the same region. As such, it is possible that the ages of the younger regions are in fact  significantly larger, as was proposed by \citet{Bell_2013}; this is further supported in a different aspect considering the need to account for magnetic fields in low mass stars by \citet{Feiden_2016}. Consequently, such arguments would extend the disk lifetimes even further for both disks in low UV regions (our sample) as well as disk lifetimes of UV-dominated regions beyond the presented results from Figure~\ref{fig:diskfrac}. 
 
 Lastly, it is becoming apparent that within young star-forming regions, the star formation process is likely spread out over a period of time \citep[e.g.,][]{Walawender_2008, Krumholz_2014, Soderblom_2014}. This results in sub-groups within a region which could be of different age \citep[e.g.][]{Galli_2019, Esplin_2020, Squicciarini_2021}. Such distinctions require further analysis of each individual region to properly parse out each sub-group and define their ages consistently. We check whether splitting up a region, e.g., Ophiuchus, impacts the disk lifetime. \citet{Esplin_2020} identify six sub-groups in Ophiuchus and describe their ages in relation to Upper Sco. When Ophiuchus is split into these sub-groups, we find that the disk lifetimes results vary insignificantly, the lifetimes are within the uncertainties presented in Figure~\ref{fig:diskfrac}. Doing a complete sub-group analysis for each region is beyond the scope of this work and may not significantly impact the disk lifetime. Rather, we have presented a re-evaluation of the disk lifetime for low-UV dominated regions without separating subgroups to allow for consistency and proper comparison between the different regions.
 
\section{Summary and conclusions}\label{sec:summary}

 We have selected Class II and III disks (primordial and evolved) using infrared criteria in 11 nearby star-forming regions for which the membership is well constrained mainly using \textit{Gaia} observations. For the Class III disks in our sample, we collected available literature and ALMA archival sub-mm/mm observations and mass accretion rates to study the evolutionary trends across different protoplanetary disk phases.
 
\begin{enumerate}[noitemsep]
    \item We present a new view of the disk evolution process, where the rapid decrease in dust mass in the bulk of the Class II disks is caused by radial drift, whereas structured disks are not affected. This can explain the low dust masses (mean: $0.29~M_{\oplus}$) that are measured in the Class III disks. 
    \item Structured disks stand out in the evolutionary process as they retain high $M_{\text{dust}}$ while they undergo a gradual $L_{\text{fract}}$ decrease. Structured protoplanetary disks are the potential parent population to cold debris disks where we subsequently observe significant dust belts at large radii, 
    as the dust does not drift inwards in these disks.
    \item In contrast, the majority of Class II disks appear to be dominated by radial drift, which likely result in near diskless stars / diskless stars at current debris disk observation sensitivities.
    \item We present updated characteristic timescales of ${\sim}8$~Myr for the disk fraction evolution in time, which is 2-3 times larger than previous estimates, although the exact value remains debatable considering the uncertainties in the age determinations of young clusters. The longer lifetime is consistent with slow dissipation and low viscosities as explored by \citet{Sellek_2020}.
\end{enumerate}

 As we continue to develop a better and more thorough understanding of disk evolution, we pose some new questions: How low can Class III mm-dust masses be, and do they display the same distribution (across 3 orders of magnitude) as Class II disks? Are they under-represented due to sensitivity as proposed by \citet{Luppe_2020} or are most M-star protoplanetary disks so significantly radial drift dominated that these result in near diskless stars? At what moment does the UV-switch take place, overtaking the accretion rate and rapidly clearing out the gas? Are the Class III disk detections in our sample structured disks that have decreased the impacts of radial drift and are precursors to the debris disks? Future higher sensitivity observations of Class III disks will answer some of these questions and help us further develop our understanding of disk evolution. The proposed disk lifetimes are dependent on star-forming regions' membership and age determinations, therefore future developments to consistently address region sub-groups, membership, and ages is a key future development.

\acknowledgements{We thank the referees for their constructive reports and useful suggestions that have significantly improved our manuscript. We would like to thank Jonathan Williams, Bruno Mer{\'\i}n, Richard Booth, and Joshua Lovell for useful discussions. We would also like to thank G. M. Kennedy for providing supplementary debris disk target data from the SONS survey and constructive discussions, K. L. Luhman for the Lupus star-forming region tables and discussion, and P. Galli for the Chamaeleon I \& II tables.}
N.M. acknowledges support from the Banting Postdoctoral Fellowships program, administered by the Government of Canada.
This work has made use of data from the European Space Agency (ESA) mission {\it Gaia} (\url{https://www.cosmos.esa.int/gaia}), processed by the {\it Gaia} Data Processing and Analysis Consortium (DPAC, \url{https://www.cosmos.esa.int/web/gaia/dpac/consortium}). Funding for the DPAC has been provided by national institutions, in particular the institutions participating in the {\it Gaia} Multilateral Agreement.
ALMA is a partnership of ESO (representing its member states), NSF (USA) and NINS (Japan), together with NRC (Canada) and NSC and ASIAA (Taiwan) and KASI (Republic of Korea), in cooperation with the Republic of Chile. The Joint ALMA Observatory is operated by ESO, AUI/ NRAO and NAOJ. This paper makes use of the following ALMA data: 2011.0.00733.S, 2012.1.00313.S, 2013.1.01075.S, 2016.1.01511.S, 2017.1.01729.S, 2017.1.01627.S, and 2018.1.00564.S.
We acknowledge the use of the ARCADE (ALMA Reduction in the CANFAR Data Environment) science platform. ARCADE is a ALMA Cycle 7 development study with support from the National Radio Astronomy Observatory, the North American ALMA Science center, and the National Research center of Canada.
This research has made use of the SIMBAD database, operated at CDS, Strasbourg, France.
This publication makes use of data products from the Wide-field Infrared Survey Explorer, which is a joint project of the University of California, Los Angeles, and the Jet Propulsion Laboratory/California Institute of Technology, funded by the National Aeronautics and Space Administration.

\bibliography{references}{}
\bibliographystyle{aasjournal}

\begin{appendix}

\section{Low-mass nearby star-forming regions}\label{appendix:regions}

 \paragraph{Ophiuchus} 
 The Ophiuchus star-forming region was re-characterized and constrained by \citet{Esplin_2020} to having 373 candidate members based on \textit{Gaia} selection and color-magnitude diagrams based on photometry. The high extinction present toward of Ophiuchus and the region's expected relative youth \citep{Wilking_2008}, indicate that we expect many Class I+F targets which \textit{Gaia} cannot observe \citep{Canovas_2019}. To address this issue, we add 47 Class I/F targets obtained from \citet{Evans_2009} and \citet{Dunham_2015} that were not observed by Gaia nor included by \citet{Esplin_2020} and thus cannot be ruled out as not being members. This produces a sample of 420 Ophiuchus members for which we collect IR data for 273 members from \textit{Spitzer} \citep{Evans_2009, Dunham_2015} and 121 other members from \textit{WISE}. The IR data distribution of the 394 members is of 69\% from \textit{Spitzer} and 31\% from \textit{WISE}.
 
 \paragraph{Taurus}
 The Taurus star-forming region was updated by \citet{Esplin_2019} to having 591 members using \textit{Gaia} data. The sample is assumed to be near complete according to \citet{Esplin_2019} as they use both \textit{Gaia} when available and complement sources without such observations with colour-magnitude diagrams and proper motions measured with multiple wide-filed optical and IR surveys. The IR data is provided by \citet{Esplin_2019} for most members; in the context of our analysis, there are 467 targets with sufficient data to categorize the targets. The IR data distribution is of 76\% from \textit{Spitzer} and 24\% from \textit{WISE}.
 
 \paragraph{Chamaeleon I}
 The Chamaeleon I star-forming region, also referred to as Cham I in the text, was re-characterized by \citet{Galli_2021} to having 188 members using \textit{Gaia} data. Although extinction towards Chamaeleon as a whole is expected to be moderate in comparison to Ophiuchus \citep{Cambresy_1999}, we supplement the sample with 15 Class I/F targets identified by \citet{Luhman_2008_cham} and \citet{Dunham_2015} which are lacking \textit{Gaia} observations. This produces a sample of 202 Chamaeleon I members for which we collect IR data for 67 members from \textit{Spitzer} \citep{Luhman_2008_cham,Dunham_2015} and 116 other members from \textit{WISE}. The IR data distribution of the 183 members is of 37\% from \textit{Spitzer} and 63\% from \textit{WISE}.
 
 \paragraph{Chamaeleon II}
 The Chamaeleon II star-forming region, also referred to as Cham II in the text, was re-characterized by the same paper as Chamaeleon I, \citet{Galli_2021}, to having 41 members using \textit{Gaia} data. We supplement the sample with 6 Class I/F targets identified by \citet{Alcala_2008} and \citet{Evans_2009} which are lacking \textit{Gaia} observations. This produces a sample of 47 Chamaeleon II members for which we collect IR data for 18 members from \textit{Spitzer} \citep{Alcala_2008,Evans_2009} and 23 other members from \textit{WISE}. The IR data distribution of the 41 members is of 44\% from \textit{Spitzer} and 56\% from \textit{WISE}.
 
 \paragraph{IC 348}
 The IC 348 star-forming region has not yet been re-characterized with \textit{Gaia} data. So, we proceed to do a simple \textit{Gaia} distance-based membership selection of the IC 348 catalogs from \citet{Lada_2006} and \citet{Evans_2009}. We collect a total of 365 targets by merging both catalogs and cross-match these with \textit{Gaia} observations using similar queries as to those explained in Appendix A of \citet{Manara_2018}. Based on these we exclude any YSO with a distance (obtained by \textit{Gaia} parallax inversion) beyond the selected 200 to 500~pc bounds; IC 348 is expected to be located at a distance of about 310~pc \citep{Ruiz-Rodriguez_2018}. This leave us with 349 targets that fit a YSO and distance-criterion to be part of IC 348. We acknowledge that this is a weaker membership selection treatment compared to other star-forming regions; however, it appears that a full treatment based on more complex and complete \textit{Gaia}-based selection would not significantly influence the membership lists (Kevin Luhman, private comm.). As this sample is built on \textit{Spitzer}-based catalogs, 100\% of the IR data are from \textit{Spitzer}.
 
 \begin{figure}[!t]
   \centering
   \includegraphics[width=0.5\textwidth]{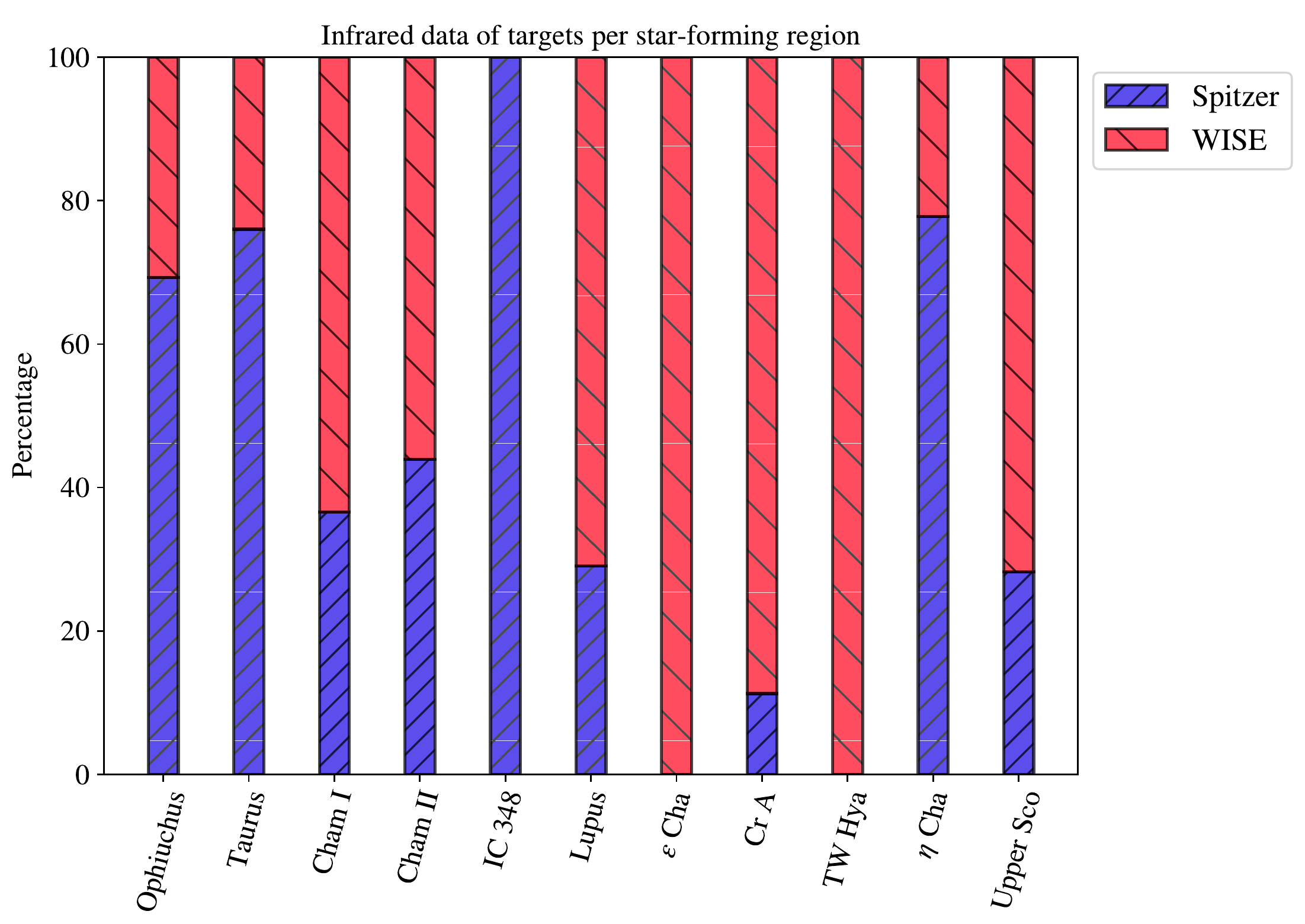}
      \caption{Bar plot of the source (\textit{Spitzer} or \textit{WISE}) of the infrared photometry per star-forming region.} \label{fig:IRsource}
 \end{figure}
 
 \paragraph{Lupus}
 The Lupus star-forming region was re-characterized by \citet{Luhman_2020_lupus} to having 214 members using \textit{Gaia} data. The sample is assumed to be near complete as \citet{Luhman_2020_lupus} uses both \textit{Gaia} when available and also relies on a careful analysis of previous IR surveys to add extra candidates that lack \textit{Gaia} observations. The \textit{WISE} IR data is provided by \citet{Luhman_2020_lupus} for most of the members but we give priority to \textit{Spitzer} data and thus collect IR data for 57 targets \citet{Merin_2008,Evans_2009,Dunham_2015} and have sufficient \textit{WISE} data for another 139 targets. We thus have a total of 196 targets The IR data distribution is of 29\% from \textit{Spitzer} and 71\% from \textit{WISE}.
 
 \paragraph{$\epsilon$ Chamaeleontis}
 The $\epsilon$ Chamaeleontis star-forming region, also referred to as $\epsilon$ Cha in the text, was well constrained and studied by \citep{Murphy_2013} using proper motions and spectroscopic data. \citet{Murphy_2013} report 35 confirmed and 6 provisional members for which we collect \textit{WISE} IR data. We find IR data for 40 targets, TYC 9414-191-1 is lacking IR observations.
 
 \paragraph{Corona Australis}
 The Corona Australis star-forming region, also referred to as Cr A in the text, was re-characterized by \citet{Galli_2020_cra} to having 313 members using \textit{Gaia} data. The extinction towards the region is expected to be variable \citep{Cambresy_1999,Alves_2014}, thus we supplement the sample with 12 Class I/F targets identified by \citet{Peterson_2011} and \citet{Dunham_2015} which are lacking \textit{Gaia} observations. For this sample of 325 members, only 275 have sufficient IR data for our analysis. We collect IR data for 31 members from \textit{Spitzer} \citep{Peterson_2011,Dunham_2015} and 244 other members from \textit{WISE}. The IR data distribution of the 275 members is of 11\% from \textit{Spitzer} and 89\% from \textit{WISE}.
 
 \paragraph{TW Hydra}
 The TW Hydra star-forming region, also referred to as TW Hya in the the text, was well constrained by \citet{Gagne_2017}. The membership is conducted using low- and high-resolution, optical and near-infrared spectroscopy, radial velocity, and Hipparcos data to ascertain the probability of membership \citep{Gagne_2017}. From the sample, we select targets that are classified as `bonafide' and `high likelihood' members resulting in a sample of 40 members. All the IR data for this sample is from \textit{WISE} and available within \citet{Gagne_2017}.
 
 \paragraph{$\eta$ Chamaeleontis}
 The $\eta$ Chamaeleontis star-forming region, also referred to as $\eta$ Cha in the text, was constrained and studied by \citep{Sicilia-Aguilar_2009} using IR spectroscopy. The sample of 18 members is observed by \textit{Spitzer} for which we collect complete IR data for 14 targets \citep{Sicilia-Aguilar_2009}; the remaining 4 have incomplete IRAC/MIPS1 data but complete WISE1-4 bands thus favoring WISE data. The IR distribution is of 78\% from \textit{Spitzer} and 22\% from \textit{WISE}.
 
 \paragraph{Upper Scorpius}
 The Upper Scorpius star-forming region, also referred to as Upper Sco in the text, was re-characterized by \citet{Luhman_2020_usco} to having 1761 members using \textit{Gaia} data. IR data in the form of both IRAC/MIPS1 and WISE bands are provided by \citet{Luhman_2020_usco} for most of the members (1712). There are 483 targets with sufficient \textit{Spitzer} data which we preferentially use and have sufficient \textit{WISE} data for another 1229 targets. The IR data distribution is of 28\% from \textit{Spitzer} and 72\% from \textit{WISE}.
 
 \vspace{5mm}
 
 The ratio of IR data obtained from \textit{Spitzer} to \textit{WISE} is plotted as a bar graph per star-forming region in Figure~\ref{fig:IRsource}. We show this to illustrate the origin of the IR photometry that is used in the classification outlined below in Section~\ref{sec:classification}. Note, when a target has been previously identified as a YSO in membership studies but is lacking complete \textit{Gaia} data, the average distance of the expected star-forming region is used and the target is still considered to be a member.

\section{Fractional disk luminosity SED example}\label{appendix:Lfract_SED}
\begin{figure*}[!ht]
 \gridline{\fig{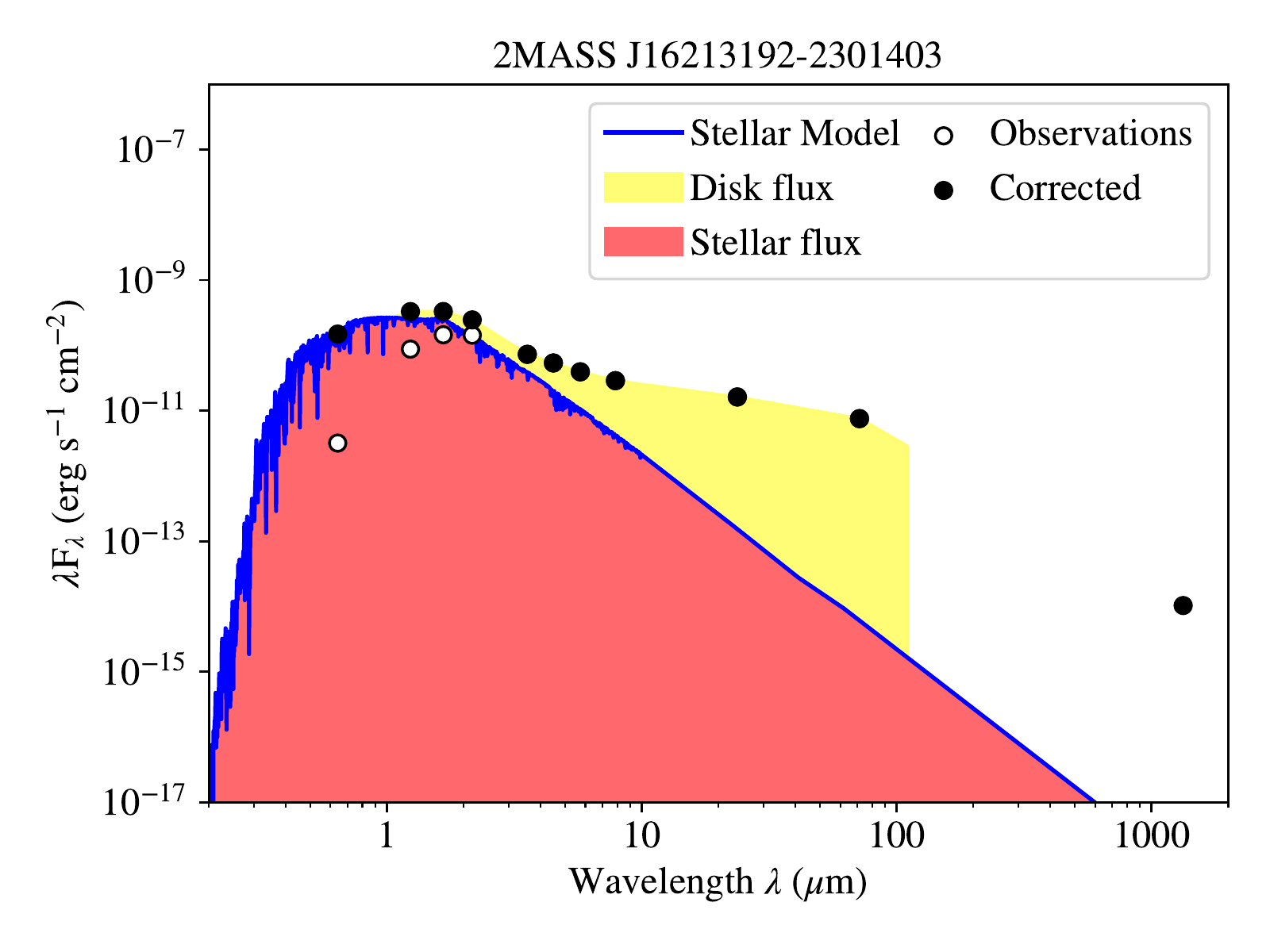}{0.5\textwidth}{(a)}
           \fig{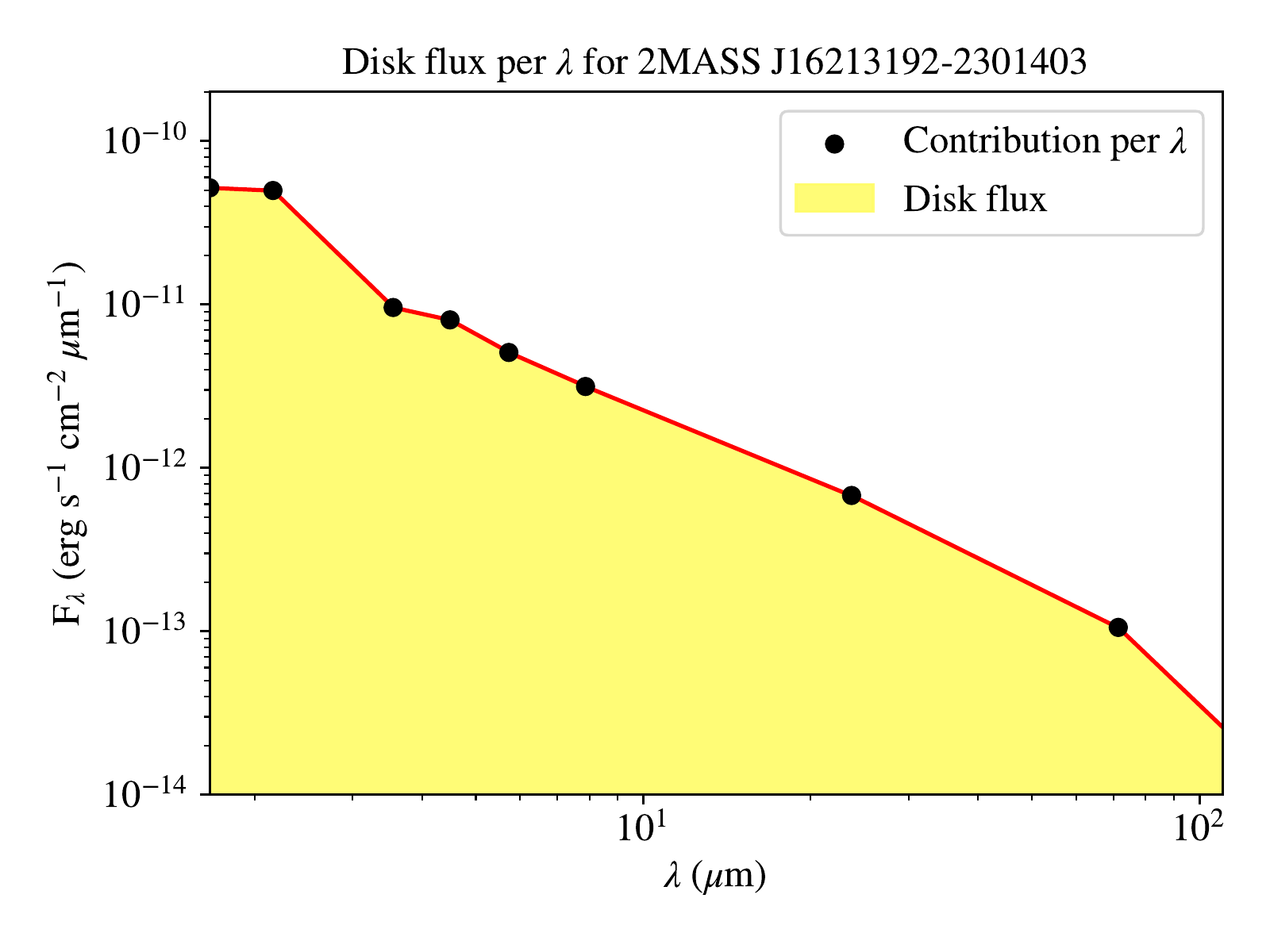}{0.5\textwidth}{(b)}}
 \gridline{\fig{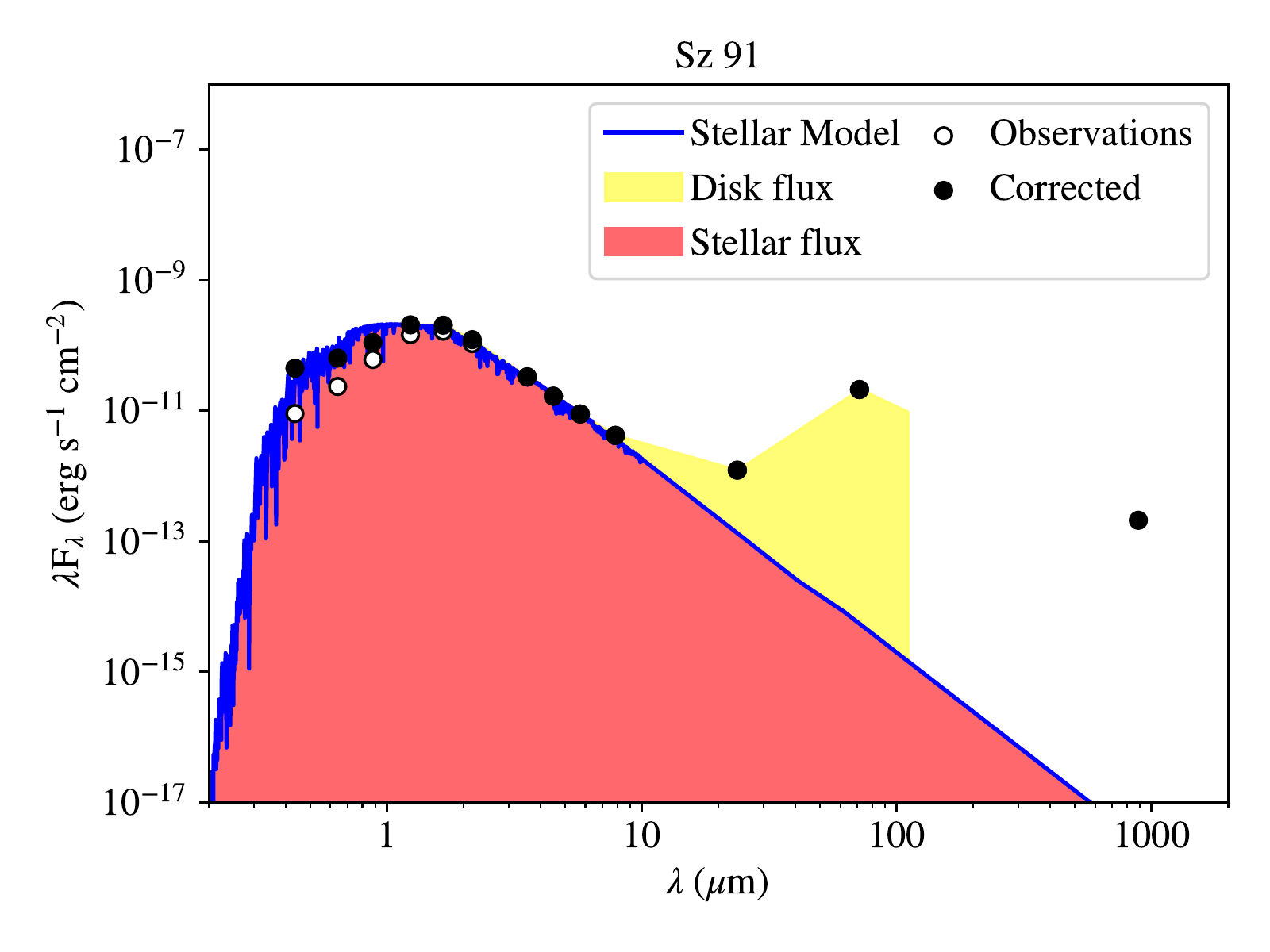}{0.5\textwidth}{(c)}
           \fig{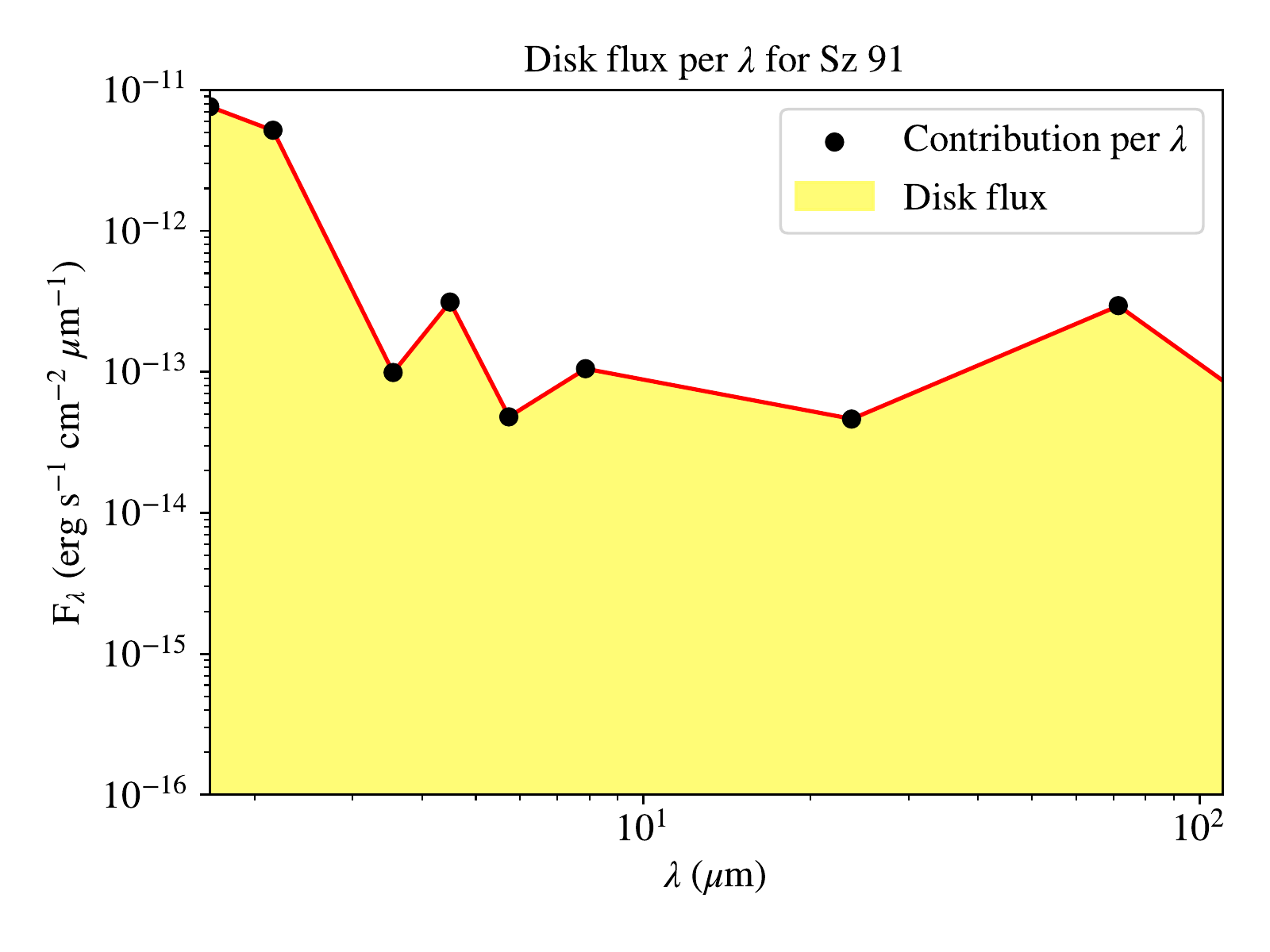}{0.5\textwidth}{(d)}}
 \caption{{\bf Left column} SED plots for a Class II (a), Sz 91 (c), and Class III (e) object. The shaded yellow demonstrates the disk flux contribution and the shaded red is the expected stellar luminosity. $L_{\text{fract}}$ is calculated from dividing the calculated disk flux by the literature stellar luminosity. {\bf Right column} Disk flux contributions per wavelength. {\bf (a):} Class II disk in Upper Sco. {\bf (b):} Disk flux for the Class II disk. {\bf (c):} Sz~91 in Lupus, a Class III disk according to the $\alpha_{\mathrm{Lada}}$ classification scheme for which there is a detected and evident 70~$\mu$m excess. {\bf (d):} Disk flux contributions for Sz~91. Although there is a significant 70~$\mu$m excess, its particular contribution to the disk flux, and consequently the $L_{\text{fract}}$ parameter is low. \label{fig:Lfract}}
\end{figure*}

\begin{figure*}[!ht]
 \gridline{\fig{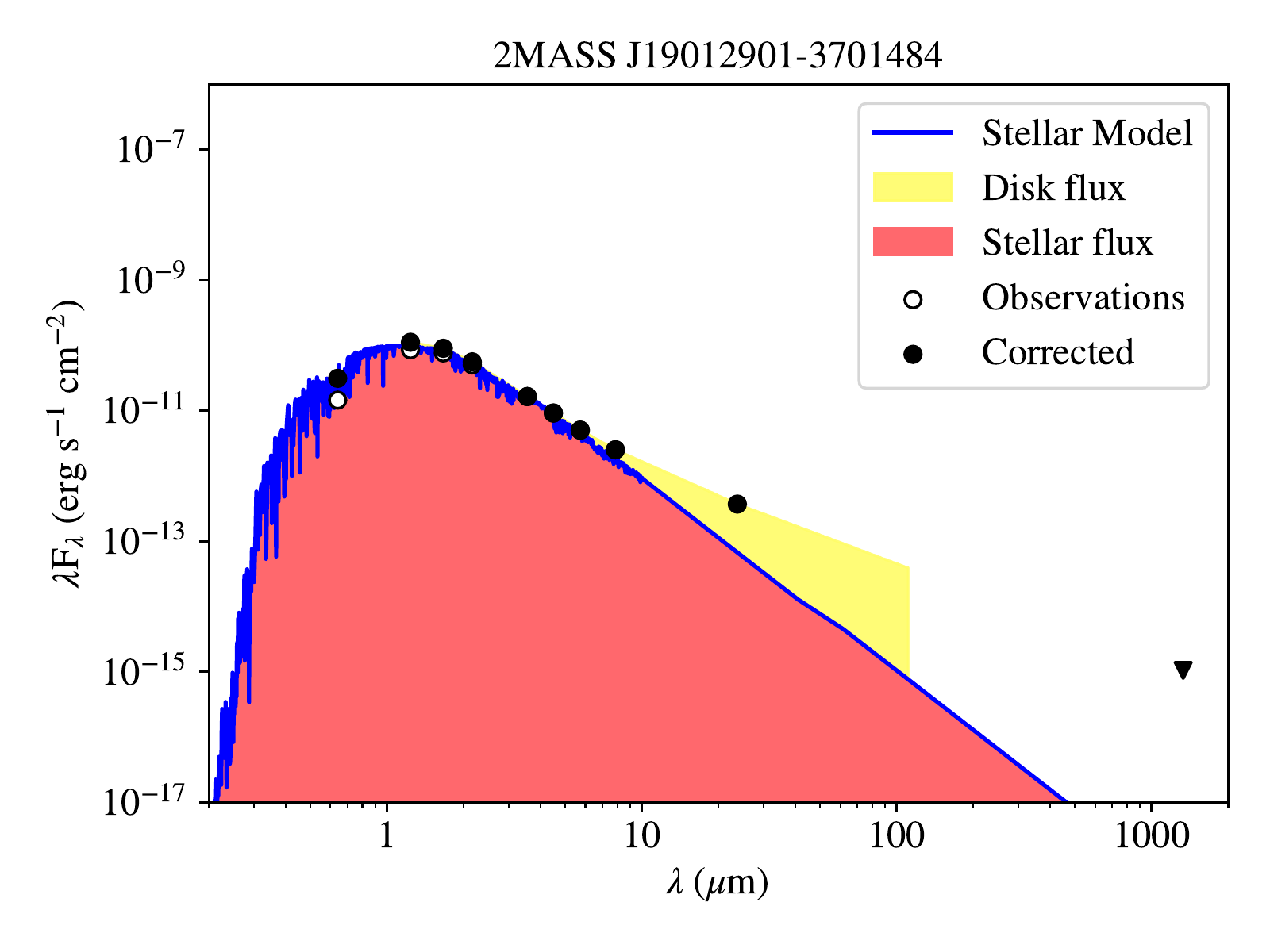}{0.5\textwidth}{(a)}
          \fig{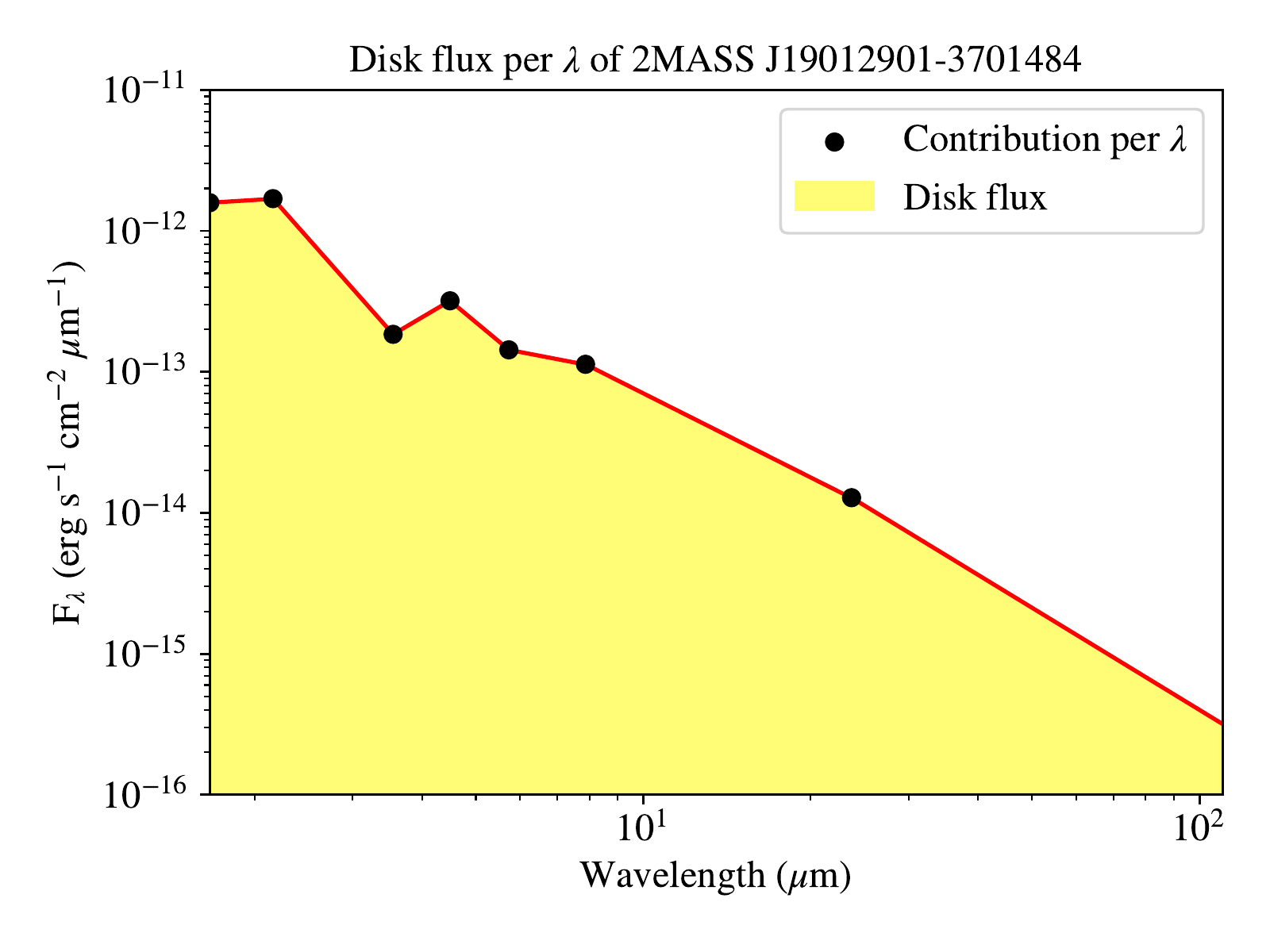}{0.5\textwidth}{(b)}
          }
 \gridline{\fig{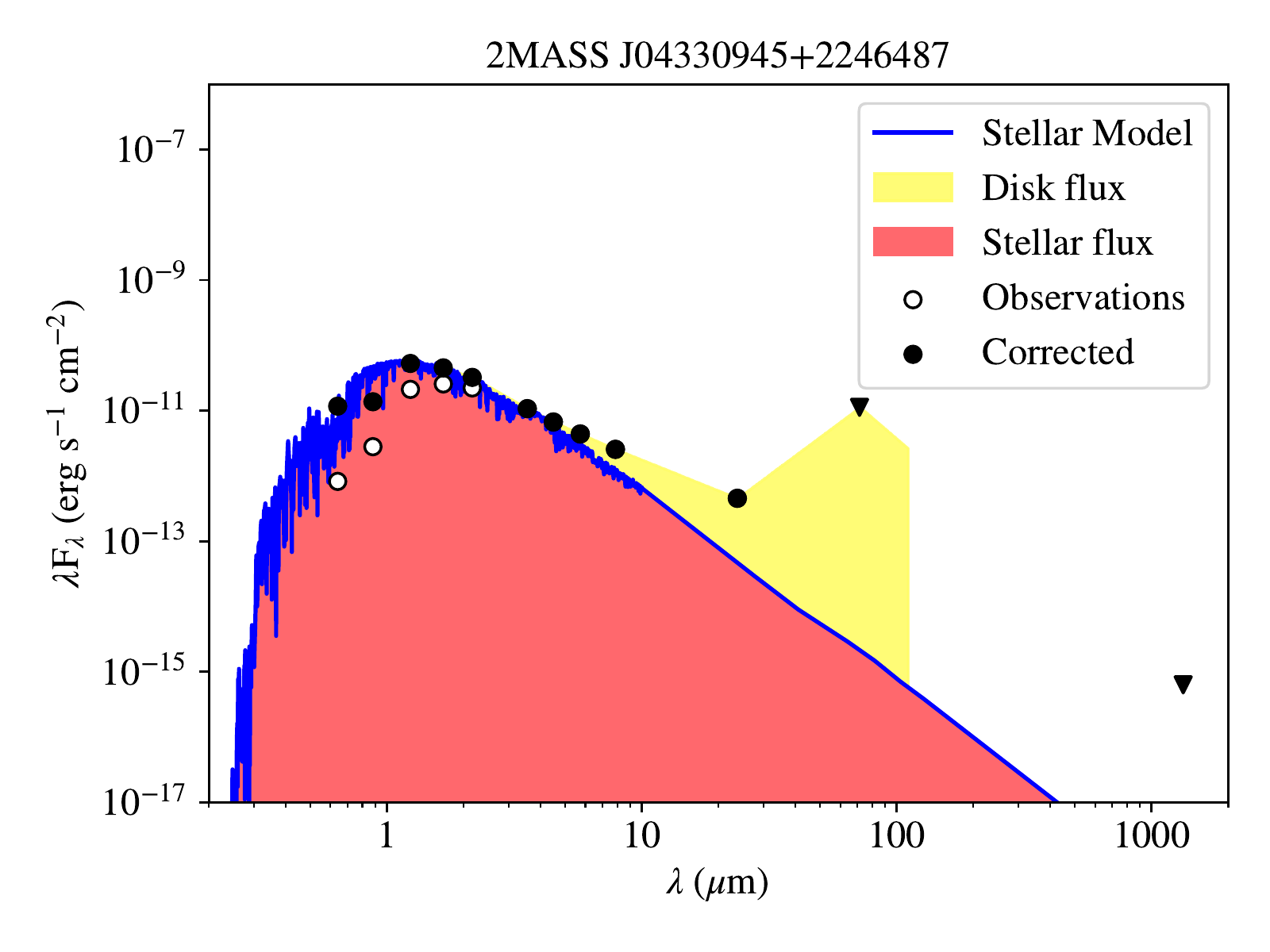}{0.5\textwidth}{(c)}
          \fig{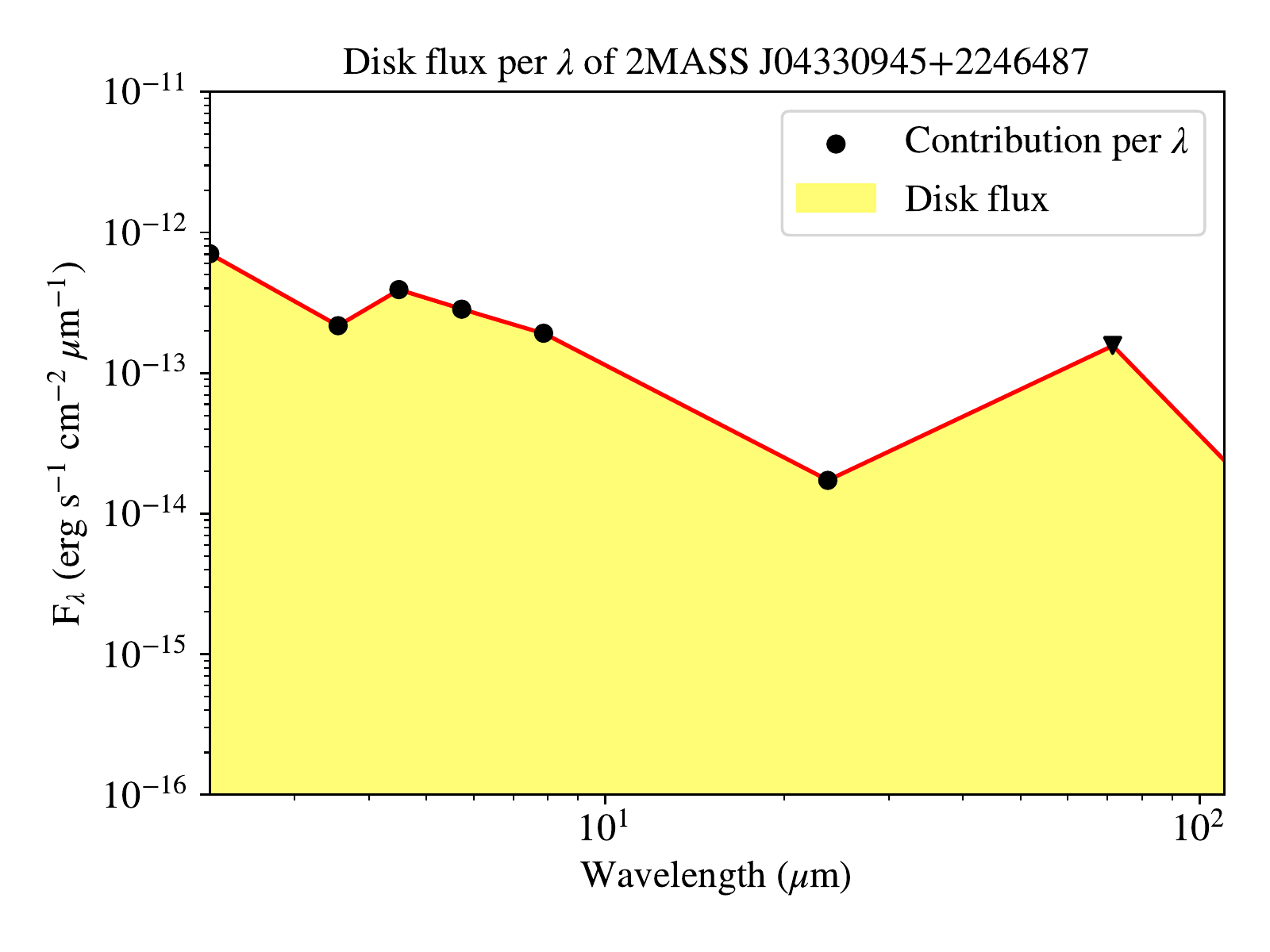}{0.5\textwidth}{(d)}
          }
 \caption{Same SED plots as Figure~\ref{fig:Lfract}
 {\bf (a):} Class III disk in Cr A. {\bf (b):} Disk flux contribution for the Class III disk. {\bf (c):} Class III disk with a 70~$\mu$m upper limit in Taurus. {\bf (d):} Disk flux contribution for the Class III disk. Even though there is a large upper limit for the evident 70~$\mu$m, it's impact on $L_{\text{fract}}$ is low changing it from $1.53\cdot10^{-2}$ to $<1.53\cdot10^{-1}$ which is an uncertain upper limit. \label{fig:Lfract2}}
\end{figure*}

 The manner by which $L_{\text{fract}}$ is calculated is illustrated in Figures~\ref{fig:Lfract} and~\ref{fig:Lfract2}. In both Figures, the left column of panels show the stellar flux and disk flux. The right column of panels show the disk flux contributions per wavelength between 1.66~$\mu$m and the millimeter flux, which are interpolated in the far infrared regime. The disk flux between 1.66~$\mu$m at 110~$\mu$m provides a $\mu$m-sized dust luminosity emanating from the disk. The shaded yellow area thus represents the disk flux and once divided by the stellar flux yields $L_{\text{fract}}$. The 70 $\mu$m \textit{Spitzer} MIPS2 flux is undetected in most Class III targets (81/85). When we consider the upper limits for the 70~$\mu$m data point (available for 15/85 Class III targets), the calculated $L_{\text{fract}}$ value is at most a factor of 2 larger which makes its inclusion both uncertain and largely insignificant compared to the larger spread in $L_{\text{fract}}$ values between Class II, III and debris disk objects. For consistency across the Class III disk sample, we have chosen to exclude the upper limits and non-detections from MIPS2 fluxes in $L_{\text{fract}}$. Furthermore, the MIPS2 70~$\mu$m flux is likely to be contaminated by nearby sources when observing YSOs in clusters, due to the large beam size; in considering IRAS-60~$\mu$m and IRAS-100~$\mu$m fluxes, the problem would be further exacerbated by the larger beam sizes and thus these are not used.

\section{Independence of $L_{\text{fract}}$ and $\dot{M}_{\text{acc}}$ from $L_{*}$}\label{appendix:Lfract_Lstar_Macc}
\begin{figure*}[!ht]
 \gridline{\fig{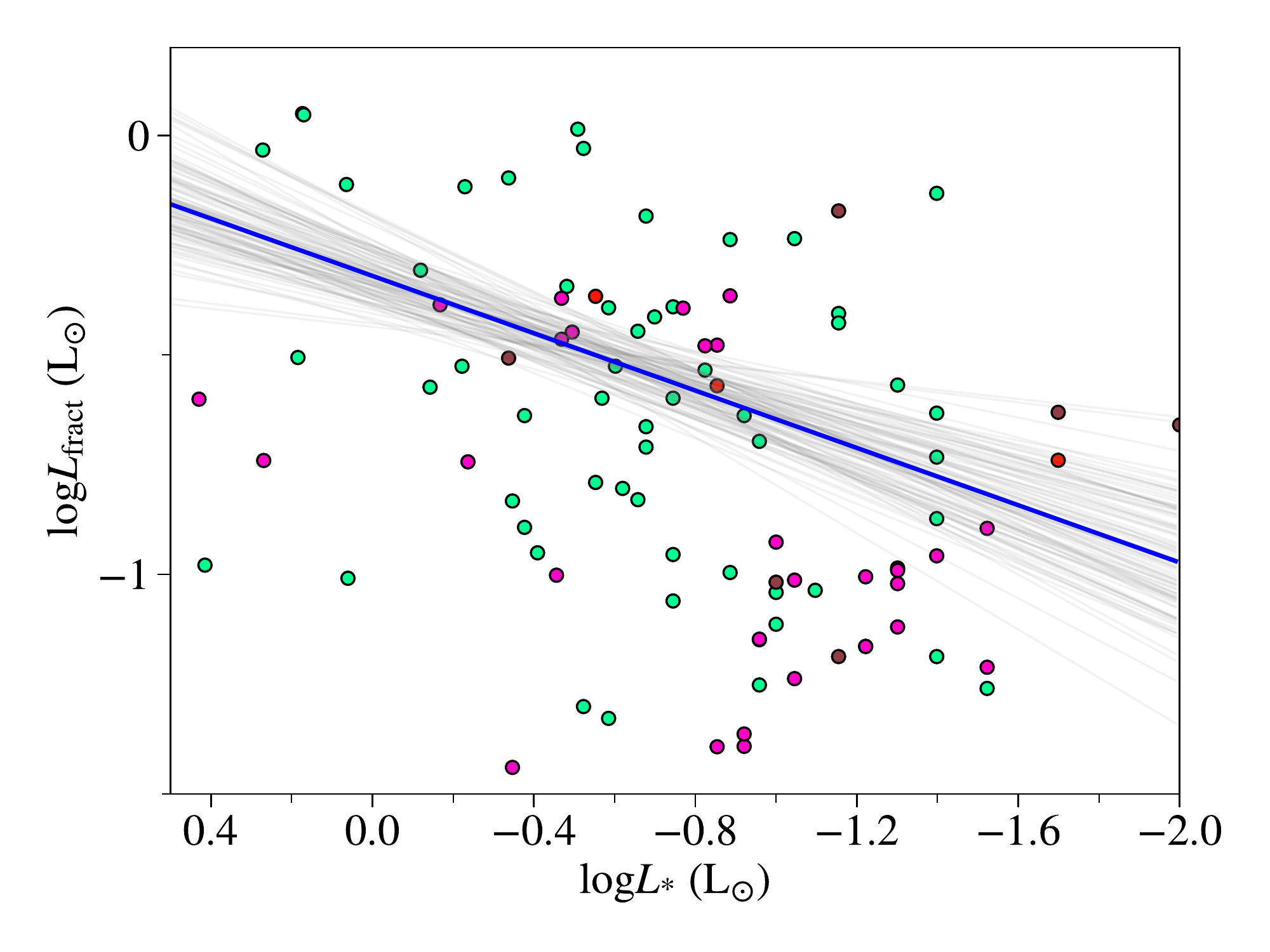}{0.5\textwidth}{(a)}
          \fig{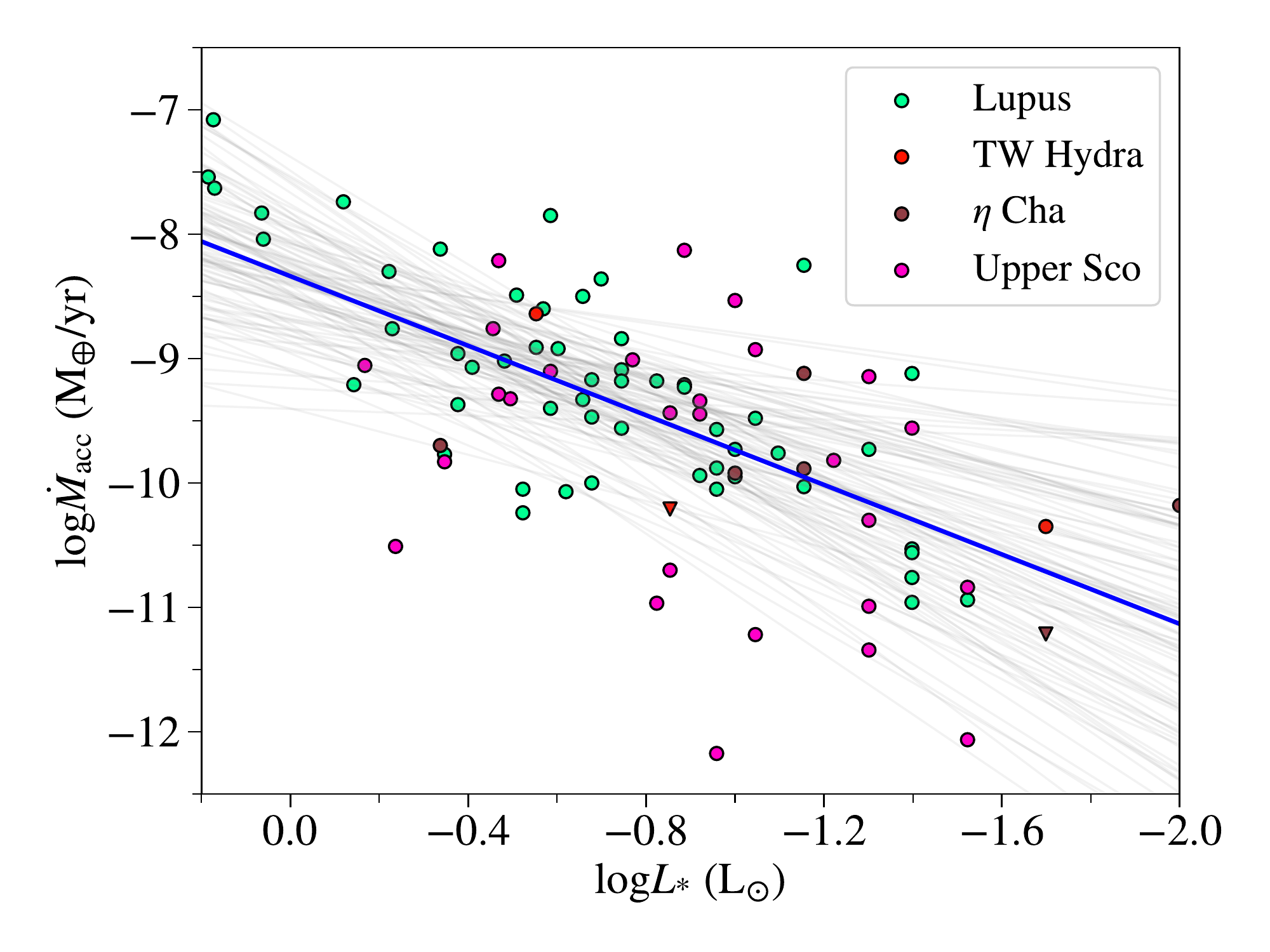}{0.5\textwidth}{(b)}
          }
 \caption{Both of these plots include Class II objects which have $\dot{M}_{\text{acc}}$ measurements for which we conduct a brief analysis as to the parameter dependencies. {\bf Left:} $L_{\text{fract}}$ vs $L_{*}$ relationship for which we find a mediocre correlation $r_{\rm corr}=0.53\pm0.12$. {\bf Right:}  $\dot{M}_{\text{acc}}$ vs vs $L_{*}$ relationship for which we find a stronger correlation $r_{\rm corr}=0.81\pm0.2$. The legend in the right panel (b) applies to both plots. \label{fig:Lfract_Lstar_Macc}}
\end{figure*}

We compute the correlation between $L_{\text{fract}}$ and $L_{*}$ to investigate and show that the derived $L_{\text{fract}}$ - $\dot{M}_{\text{acc}}$ correlation is significant, shown in Figure \ref{fig:Lfract_Lstar_Macc}. We use an uncertainty of 20\% for Class II $L_{*}$ based on similar extinction and objects as in \citet{vanderMarel_2019,Francis_2020}.

\section{SEDs for Class III objects with mm-observations}\label{appendix:SEDs}

\begin{figure}[!ht]
    \centering
    \includegraphics[width=\textwidth]{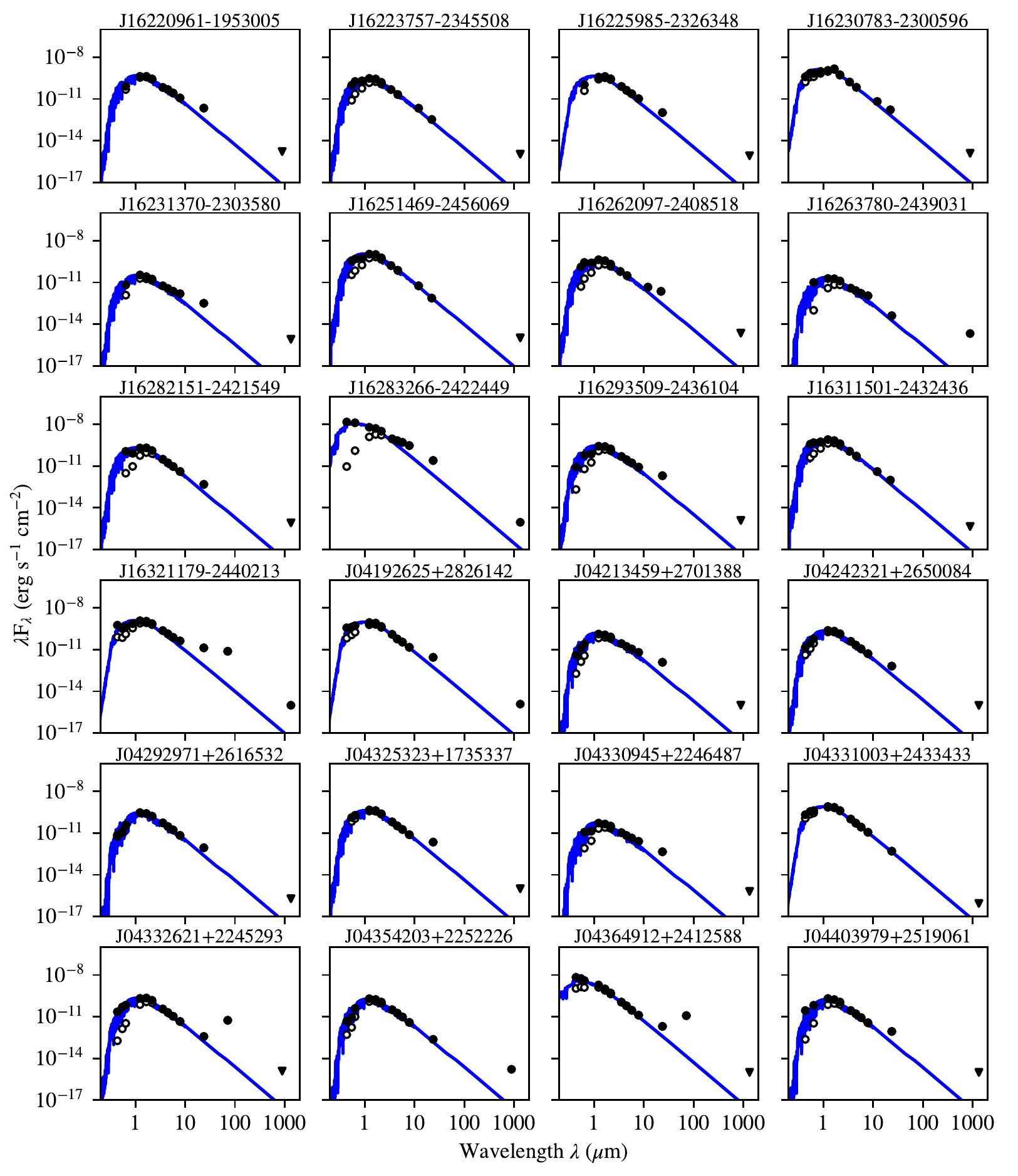}
    \captcont{Class III objects SED plots}
\end{figure}

\begin{figure}[!ht]
    \centering
    \includegraphics[width=\textwidth]{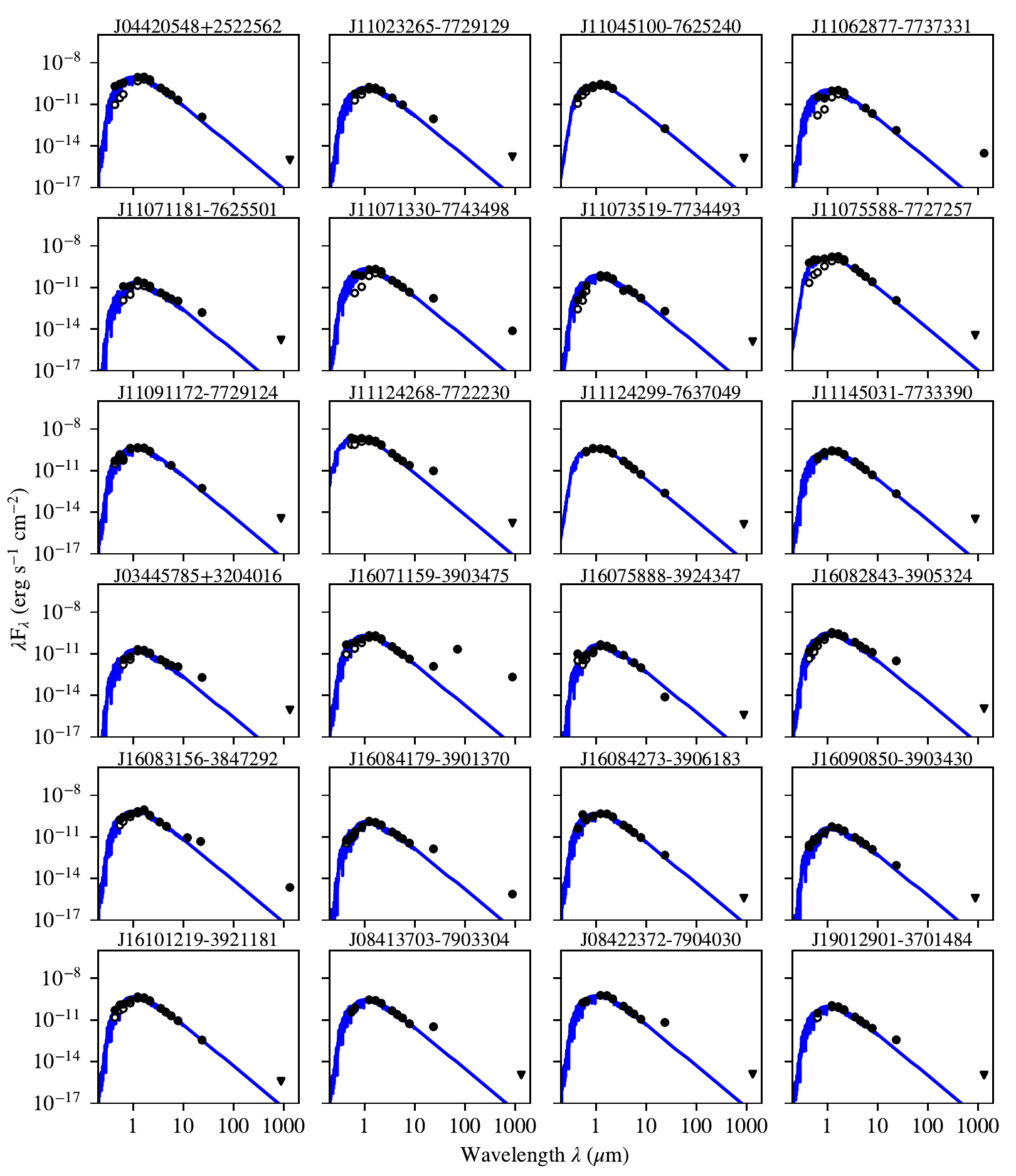}
    \captcont[]{Class III objects SED plots (cont.)}
\end{figure}

\begin{figure}[!ht]
    \centering
    \includegraphics[width=\textwidth]{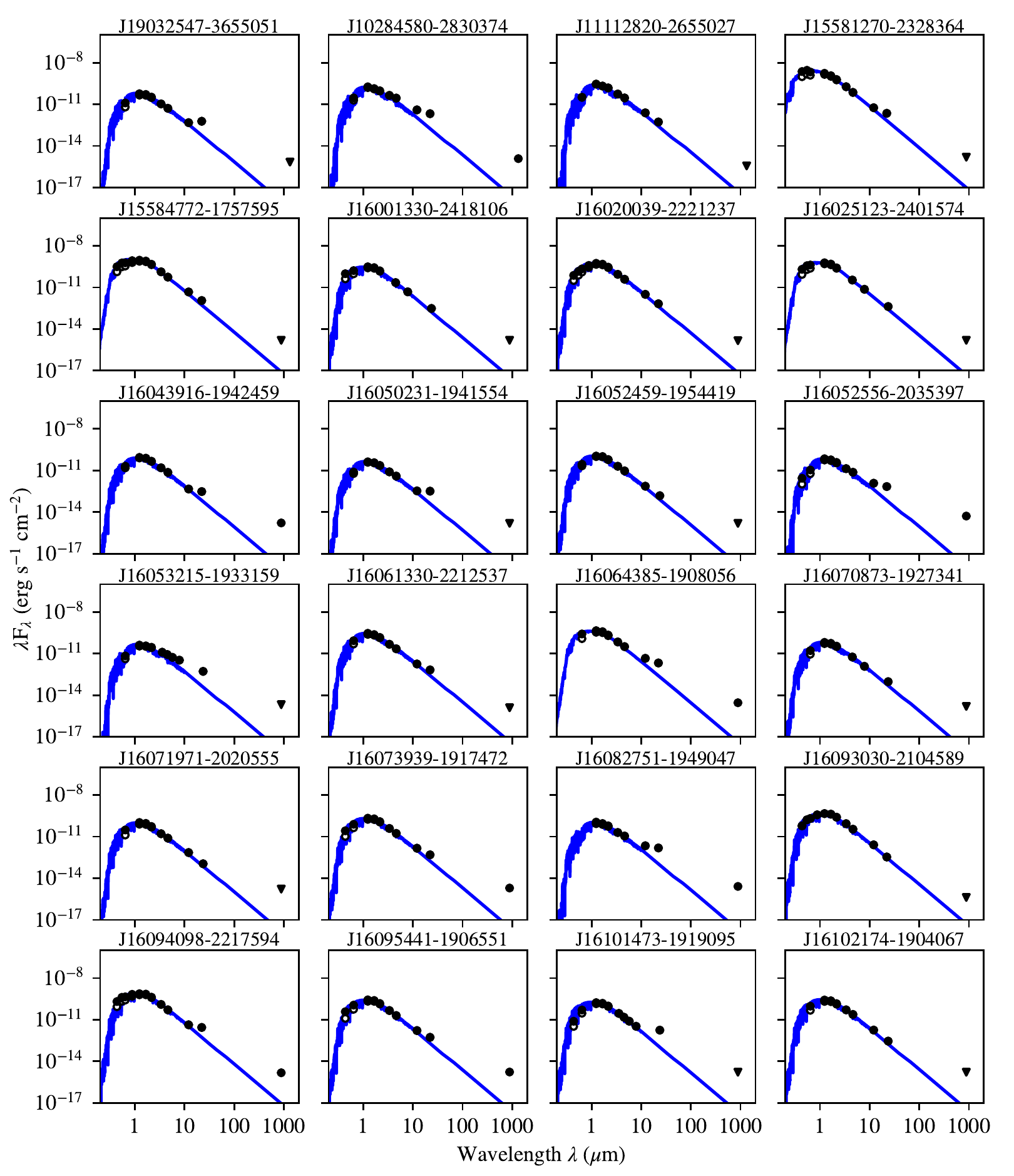}
    \captcont[]{Class III objects SED plots (cont.)}
\end{figure}

\begin{figure}[!ht]
    \centering
    \includegraphics[width=\textwidth]{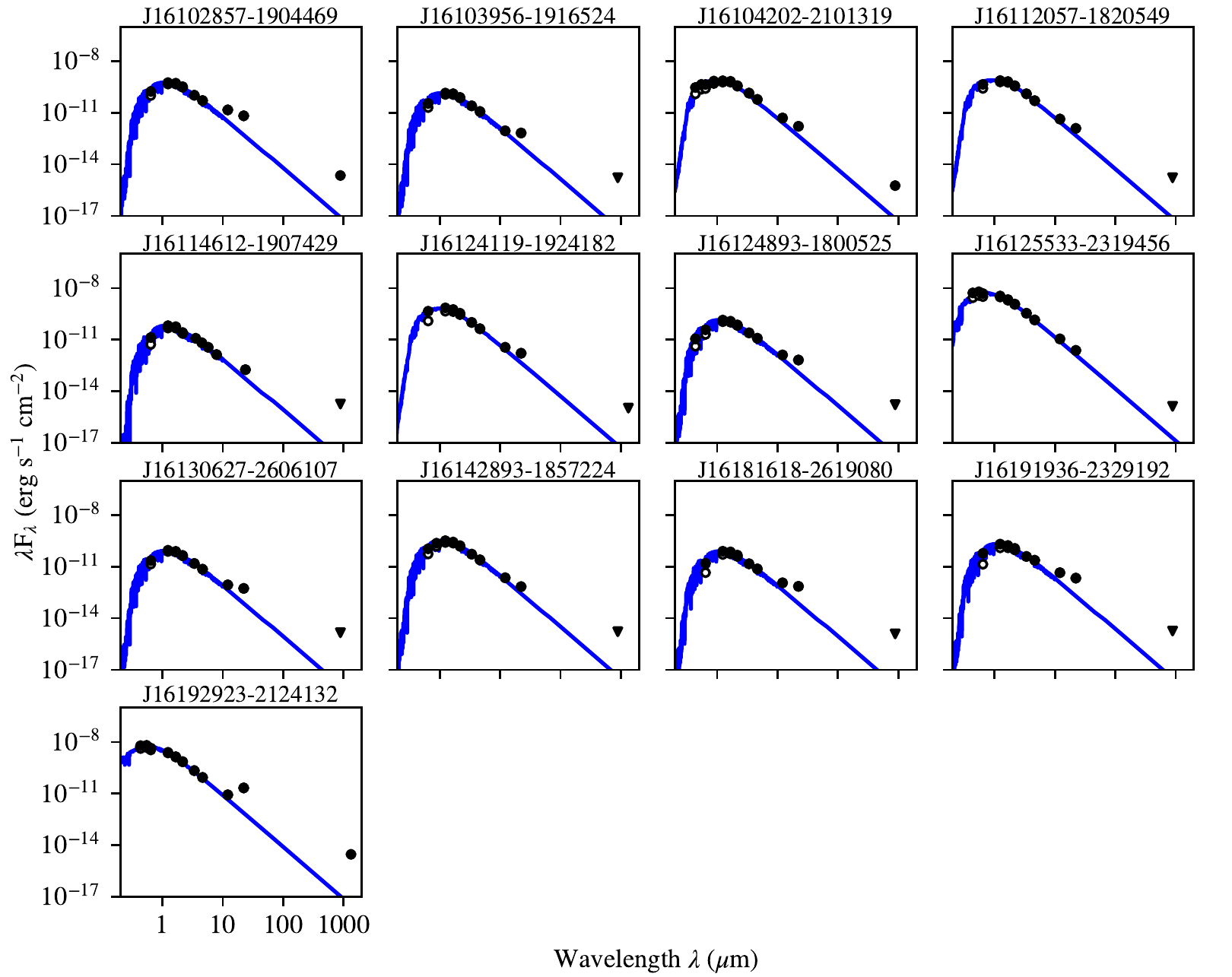}
    \captcont[]{Class III objects SED plots (cont.)}
    \label{fig:seds}
\end{figure}

\newpage
\section{Class II and III data}\label{appendix:data}

\begin{table*}[!ht]
\caption{Class III targets with millimeter-flux or with $\dot{M}_{\text{acc}}$ measurements.}
\label{tbl:mastertable}
\begin{center}
\begin{tabular}{ll}
    \hline
    Column label & Description \\
    \hline
    Region & Star-forming region object belongs to \\
    2MASS & 2MASS Point Source Catalog source name \\
    Gaia & \textit{Gaia} DR2 source name \\
    $d$ & Distance to target$^{(a)}$ based on the inversion of \textit{Gaia} parallax (pc) \\
    SpT & Spectral type \\
    ref\_SpT & Spectral type reference \\
    $A_v$ & Extinction in $V$-band (mag) \\
    ref\_$A_v$ & Extinction reference \\
    $L_*$ & Luminosity ($L_{\odot}$) \\
    ref\_$L_*$ & Stellar luminosity reference \\
    $M_*$ & Stellar mass ($M_{\odot}$) \\
    ref\_$M_*$ & Stellar mass evolutionary isochrone model used \\
    $\dot{M}_{\text{acc}}$ & Mass accretion rate ($M_{\odot}$ yr$^{-1}$) \\
    ref\_$\dot{M}_{\text{acc}}$ & Mass accretion rate reference \\
    $L_{\text{fract}}$ & Fractional disk luminosity \\
    $F_{mm}$ & ALMA millimeter flux (mJy) \\
    e\_$F_{mm}$ & Uncertainty in ALMA millimeter flux \\
    ref\_$F_{mm}$ & ALMA millimeter flux reference \\
    Band & ALMA band used for millimeter flux observations \\
    $M_{mmdust}$ & Disk dust mass ($M_{\oplus}$) \\
    e\_$M_{mmdust}$ & Uncertainty in disk dust mass \\
    Fit & Flag when $A_v$ or $L_*$ used are beyond literature reported uncertainties \\
    \hline
\end{tabular}
\end{center}
(a) When no distance is reported in the table due to a lack of \textit{Gaia} result, we adopt the average distance to the region found in Table~\ref{tbl:regiontable}.

References: 1) \citet{Pecaut_2012}; 2) \citet{Uyama_2017}; 3) \citet{Siess_2000} 4) \citet{Lieman-Sifry_2016};  5) \citet{Martin_1998}: 6) \citet{Wahhaj_2010}; 7) \citet{Baraffe_2015}; 8) \citet{Hardy_2015}; 9) \citet{Cheetham_2015}; 10) \citet{Huber_2016}; 11) This work; 12) \citet{Williams_2019}; 13) \citet{Luhman_2012}; 14) \citet{Barenfeld_2016}; 15) \citet{Dunham_2015}; 16) \citet{Erickson_2011}; 17) \citet{Prato_2003}; 18) \citet{Bowler_2017}; 19) \citet{Wu_2020}; 20) \citet{Kohn_2016}; 21) \citet{Wilking_2005}; 22) \citet{McClure_2010}; 23) \citet{Najita_2015}; 24) \citet{Bouvier_1992}; 25) \citet{Kraus_2014}; 26) \citet{Nuernberger_1998}; 27) \citet{Ribas_2017}; 28) \citet{Luhman_2004}; 29) \citet{Bulger_2014}; 30) \citet{Akeson_2019}; 31) \citet{Ward-Duong_2018}; 32) \citet{Kraus_2017}; 33) \citet{Rebull_2010}; 34) \citet{Zhang_2018}; 35) \citet{Esplin_2014}; 36) \citet{Csepany_2017}; 37) \citet{Kraus_2015}; 38) \citet{Herczeg_2014}; 39) \citet{Luhman_2009}; 40) \citet{Scelsi_2008}; 41) \citet{Cieza_2012}; 42) \citet{vandermarel_2016}; 43) \citet{Hartigan_2003}; 44) \citet{Manara_2017}; 45) \citet{Long_2018}; 46) \citet{Luhman_2007}; 47) \citet{Daemgen_2016}; 48) \citet{Frasca_2015}; 49) \citet{Daemgen_2013}; 50) \citet{Long_2017}; 51) \citet{Pascucci_2016}; 52) \citet{Lada_2006}; 53) \citet{Ruiz-Rodriguez_2018}; 54) \citet{Alcala_2017}; 55) \citet{vandermarel_2018}; 56) ; 57) \citet{Comeron_2009}; 58) \citet{Lovell_2021}; 59) \citet{Mortier_2011}; 60) \citet{Galli_2015}; 61) \citet{Manara_2013_xshooter}; 62) \citet{Merin_2008}; 63) \citet{Rugel_2018}; 64) \citet{Sicilia-Aguilar_2008}; 65) \citet{Currie_2011}; 66) \citet{Harvey_2014}; 67) \citet{Cazzoletti_2019}; 68) \citet{Luhman_2017}; 69) \citet{Liu_2015}; 70) \citet{Rodriguez_2015}; 71) \citet{Gagne_2017}; 72) \citet{Rizzuto_2015}; 73) \citet{Luhman_2020_usco}; 74) \citet{Luhman_2018_usco}; 75) \citet{Preibisch_2001}; 76) \citet{Pecaut_2016}; 77) \citet{Walter_1994}; 78) \citet{Preibisch_2002}; 79) \citet{Esplin_2018}; 80) \citet{Houk_1988}; 81) \citet{Manara_2020}; 82) \citet{Venuti_2019}; 83) \citet{Fang_2017}; 84) \citet{Romero_2012}; 85) \citet{Esplin_2020}; 86) \citet{Muzic_2012}; 87) \citet{Cieza_2010}; 88) \citet{Evans_2009}.

\end{table*} 

\begin{table*}[!ht]
\caption{Class II targets (in Lupus, $\eta$ Cha, TW Hydra, and Upper Sco) with millimeter-flux or with $\dot{M}_{\text{acc}}$ measurements used in Figures~\ref{fig:asurv},~\ref{fig:lfractmdust}, and~\ref{fig:lfracfit_lfracmacc}.}
\label{tbl:mastertable_classii}
\begin{center}
\begin{tabular}{ll}
    \hline
    Column label & Description \\
    \hline
    Region & Star-forming region object belongs to \\
    2MASS & 2MASS Point Source Catalog source name \\
    Gaia & \textit{Gaia} DR2 source name \\
    Other name & Alternate name \\
    $d$ & Distance to target$^{(a)}$ based on the inversion of \textit{Gaia} parallax (pc) \\
    SpT & Spectral type \\
    ref\_SpT & Spectral type reference \\
    $A_v$ & Extinction in $V$-band (mag) \\
    ref\_$A_v$ & Extinction reference \\
    $L_*$ & Luminosity ($L_{\odot}$) \\
    ref\_$L_*$ & Stellar luminosity reference \\
    $M_*$ & Stellar mass ($M_{\odot}$) \\
    ref\_$M_*$ & Stellar mass evolutionary isochrone model used \\
    $\dot{M}_{\text{acc}}$ & Mass accretion rate ($M_{\odot}$ yr$^{-1}$) \\
    ref\_$\dot{M}_{\text{acc}}$ & Mass accretion rate reference \\
    $L_{\text{fract}}$ & Fractional disk luminosity \\
    $F_{mm}$ & ALMA millimeter flux (mJy) \\
    e\_$F_{mm}$ & Uncertainty in ALMA millimeter flux \\
    ref\_$F_{mm}$ & ALMA millimeter flux reference \\
    Band & ALMA band used for millimeter flux observations \\
    $M_{mmdust}$ & Disk dust mass ($M_{\oplus}$) \\
    e\_$M_{mmdust}$ & Uncertainty in disk dust mass \\
    Fit & Flag when $A_v$ or $L_*$ used are beyond literature reported uncertainties \\
    Structured & Label indicating whether a disk is considered as structured based on ALMA observations. \\
    \hline
\end{tabular}
\end{center}
(a) When no distance is reported in the table due to a lack of \textit{Gaia} result, we adopt the average distance to the region found in Table~\ref{tbl:regiontable}.

References: 1) \citet{Alcala_2017}; 2) \citet{Manara_2018_stellarprop}  ; 3) \citet{Baraffe_2015}; 4) \citet{Ansdell_2018}; 5) \citet{Siess_2000}; 6) \citet{Rugel_2018}; 7) \citet{Venuti_2019}; 8) \citet{Luhman_2020_usco}; 9) \citet{Barenfeld_2016}; 10) \citet{Manara_2020}; 11) \citet{Luhman_2012}; 12) \citet{Fairlamb_2015}; 13) \citet{Andrews_2018}; 14) \citet{Rigliaco_2015}; 15) \citet{Evans_2009}; 16) \citet{Williams_2019}; 17) \citet{vandermarel_2016}; 18) \citet{Ribas_2017}; 19) Ruiz-Rodriguez et al., in prep

\end{table*} 

\end{appendix}

\end{document}